\documentclass[12pt,preprint]{aastex}

\def\ha{H$\alpha$~}
\def\hab{{H$\alpha$}}
\def\s2{[\ion{S}{2}]}
\def\o3{{[\ion{O}{3}]}}
\def\h2{{\ion{H}{2} regions}}
\def\rat{{[\ion{S}{2}]/H$\alpha$~}}
\def\ratb{{[\ion{S}{2}]/H$\alpha$}}
\def\ergs{{~erg~s$^{-1}$}}
\def\sb{{~erg~cm$^{-2}$~s$^{-1}$~arcsec$^{-2}$}}

\shorttitle{ M33 Supernova Remnants}
\shortauthors{Lee \& Lee}

\begin{document}

\title{PROPERTIES OF OPTICALLY SELECTED SUPERNOVA REMNANT CANDIDATES IN M33}
\author{Jong Hwan Lee and Myung Gyoon Lee}
\affil{Astronomy Program, Department of Physics and Astronomy, \\
Seoul National University, Seoul 151-747, Korea}
\email{leejh@astro.snu.ac.kr, mglee@astro.snu.ac.kr}

\begin{abstract}

Narrow band images covering strong emission lines are efficient to survey
supernova remnants (SNRs) in nearby galaxies.
Using the narrow band images provided by the Local Group Galaxy Survey 
we searched for SNRs in M33.  
Culling the objects with enhanced \rat and round morphology
in the continuum-subtracted \ha and \s2 images, 
we produce a list of 199 sources. 
Among them 79 are previously unknown.
Their progenitor and morphology types were classified.
A majority of the sample (170 objects) are likely remnants of core-collapse SNe, 
and 29 are remnants of Type Ia SNe. 
The cumulative size distribution of these objects is found to be similar
to that of the M31 remnants  derived in the similar way.
We obtain a power law slope, $\alpha = 2.38\pm0.05$.
Thus a majority of the sources are considered to be in the Sedov-Taylor phase,
consistent with previous findings.
The histogram of the emission line ratio (\ratb) of the remnants
has two concentrations at \rat $\sim$ 0.55 and $\sim$0.8, as in M31. 
Interestingly $L_{\rm X}$ (and $L_{\rm 20cm}$) of the compact center-bright objects 
are correlated with their optical luminosity.
The remnants with X-ray emission have brighter optical surface brightness
and smaller diameters than those without X-ray emission. 

\end{abstract}


\section{INTRODUCTION}

Supernova remnants (SNRs) are a unique resource that enables 
to investigate the final stage of the evolution of massive stars or binary stars 
interacting with the surrounding interstellar medium (ISM). 
Studies of individual SNRs in the Milky Way (MW) 
are important for understanding the detailed interaction between SNRs and the ISM. 
However, determining the global properties of the MW SNRs is hard  
because of uncertain distances to individual remnants and high extinction 
in the Galactic plane where most of the SNRs are found. 
Extragalactic SNRs do not suffer from these problems 
so that they are an ideal target for investigating statistically 
their global properties and evolution as well as their relationship to the ISM. 
They provide a clue to understand the star formation history and
chemical evolution of galaxies.
                                  
Extragalactic SNRs have been identified mostly at three bands:
optical, radio and X-ray wavelengths. 
A majority of the known SNRs in nearby galaxies were detected by optical surveys
\citep{mag95,mat97a,mat97b,pan02,bla04,son09,dop10,fra12,bla12,leo13}.
We have been carrying a project to detect SNRs in nearby galaxies. 
We presented a result of the first target, M31 \citep[henceforth L14]{lee14}. 
This paper is the second of the series, presenting the results for M33. 
M33 is an ideal galaxy for SNR surveys, because it is nearby \citep[$\sim$800 kpc,][]{lee02}, 
relatively face-on \citep[$i$ = $56^{\circ}$,][]{reg94}, and a late-type spiral (SA(s)cd) galaxy. 

\citet{dod78} first detected three SNRs using optical images in M33.
Subsequent optical studies have increased the number of M33 SNRs 
to 137 \citep[henceforth L10]{lon10}. 
\citet{lon90} showed that the cumulative size distribution ($CSD$) of 50 SNRs in M33.
It follows a power law, having an index $\alpha = 2.1$. 
They estimated a SN rate to be one per 26 $-$ 300 yr from the number of SNRs.
\citet[henceforth G98]{gor98} identified 98 SNRs, of which 53 are previously unknown.
They presented a $CSD$ for the 98 SNRs 
and showed that the Sedov-Taylor (ST) expansion model is a better fit for the distribution
than the free expansion model. 
The ST expansion model is used to derive a SN rate in M33 to be one SN every 360 yr.
However, they used narrow images obtained under poor seeing ranging from 2\arcsec\ to 2\farcs4 
(which corresponds to 8-10 pc at a distance of M33). 
They could not measure the accurate size of SNRs
and could not distinguish SNRs on the outskirts of SF regions. 

L10 presented a list of 137 SNRs derived from two emission-line surveys
using Local Group Galaxy Survey (LGGS) data taken at the KPNO 4m telescope \citep{mas06,mas07} 
and observational data obtained from 0.6 m Burrell Schmidt telescope at KPNO. 
They showed that 82 of 137 SNRs are matched with X-ray objects 
derived from the ChASeM33 survey using $Chandra$. 
However, they missed a significant fraction of SNRs 
with low optical surface brightness ($SB$)($< 10^{-16}$ \sb).
They mainly discussed X-ray properties of the SNRs in M33.  
Recently, \citet{asv14} estimated a SN rate of M33 to be one SN every 140-150 yr,
using X-ray data of the M33 SNRs and physical models at various conditions of the ambient ISM.

Thus we carry out a new optical search of SNRs in M33 utilizing the images taken in the LGGS. 
Using these data, we find new M33 remnants, especially those with low $SB$, 
concentrating on their optical properties. 

\section{DATA REDUCTION} 

\subsection{Identification Method}

We utilized the \ha and \s2 images in the LGGS to find new M33 remnants.
The LGGS covered three 36\arcmin\ square fields of M33. 
We subtracted continuum sources from the narrow band images using the R-band images.
We smoothed the images with better seeing to match 
the point-spread-function in the image 
with worse seeing, using IRAF task $psfmatch$. 
We then scaled and subtracted the resulting continuum images 
from narrow band images. 

We selected the M33 remnants considering three criteria:
emission line ratio (\ratb), the morphological structure, 
and the absence of blue stars inside the sources,
as done for M31 in L14. Details are described in L14. 
We detected objects with \rat $>$ 0.4 in emission line ratio maps, 
and selected the objects with round or shell structures in each narrow band image. 
As a result, we chose 435 sources.  

We estimated an integrated emission line ratio (i.e., the ratio of integrated \rat fluxes) 
for individual remnants to select the genuine SNRs among these candidates.  
We set an approximate circular region to cover the region with \rat $> 0.4$ 
in emission line ratio (\ratb) maps, and carried out aperture photometry of these sources. 
We determined the value of the local background for individual remnants, 
using a 100 pc $\times$ 100 pc box region that shows no emission near the SNR candidate. 
We selected the objects that have integrated \rat $>$ 0.4.
In the cases of the objects with 0.4 $<$ \rat $<$ 0.6,
we examined the existence of blue stars inside the remnants.
We excluded the sources embedding blue stars.
Then we removed the objects larger than 100 pc, 
which is too big to be SNRs \citep[L10;]{fra12}.
Finally we considered 199 objects as the probable SNRs,
including 79 new ones.

L10 provided a catalog of 137 sources based on optical and X-ray data.
We examined these objects according to our identification criteria
and considered 17 of these objects to be non-SNRs such as \h2 or superbubbles.  
Of these, seven objects do not appear to have high enough \rat to be cataloged 
as SNR candidates, and they are likely \h2. 
Nine objects have sizes with $D >$ 100 pc and 
embed many blue stars inside, so that we considered them to be superbubbles. 
L10 also suggested that these objects are possible superbubbles. 
One object (L10-35) is an oxygen-dominated SNR with little emission in \ha and \s2 (L10).
Table \ref{table1} lists the 17 objects excluded according to our criteria.
Consequently, we confirmed that 120 of the 137 known objects satisfy our identification criteria.   

\begin{figure}
   \epsscale{1.0}
   \plotone{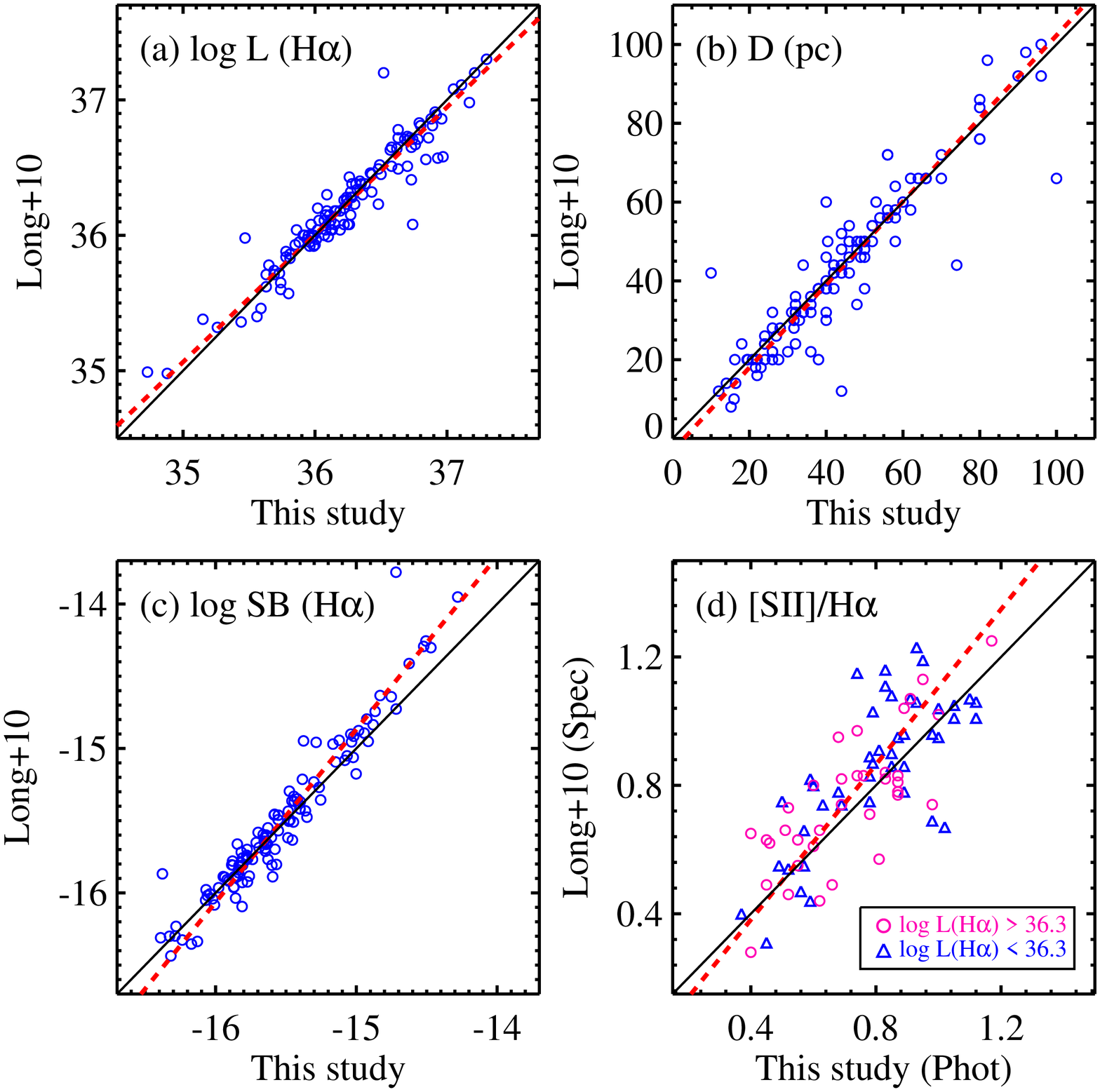}
   \caption{Comparison of (a) $L$(\hab), (b) $D$, (c) $SB$(\hab), and (d) \rat of M33 remnants 
            in this work and L10.
            Linear least-squares fit lines (dashed line)
            are close to the one-to-one relation (solid line):                       
            (a) log $L($H$\alpha)$(this work) = 0.94($\pm$0.03) $\times~$
                log $L($H$\alpha)$(L10)$~+$ 2.12($\pm$1.06) \ergs,
            (b) $D$(this work) = 1.05($\pm$0.05) $\times~D$(L10)$~-$ 3.11($\pm$1.94) pc,
            (c) log $SB($H$\alpha)$(this work) =  1.20($\pm$0.10) $\times~$
                log $SB($H$\alpha)$(L10)$~-$ 0.11($\pm$0.07) \sb, 
            and (d) \rat(this work) =  1.19($\pm$0.06) $\times~$\rat(L10)$+$ 3.07($\pm$0.89).} 
\label{comlo2}
\end{figure}

Figure \ref{comlo2} compares (a) $L$(\hab), (b) $D$, (c) $SB$(\hab), 
and (d) \rat of the remnants in this work and L10. 
Two measurements for each physical quantity of the objects 
are correlated well.
We fit the data with linear least squares fitting, obtaining 
(a) log $L($H$\alpha)$(this work) = 0.94($\pm$0.03) $\times~$
log $L($H$\alpha)$(L10)$~+$ 2.12($\pm$1.06) \ergs,
(b) $D$(this work) = 1.05($\pm$0.05) $\times~D$(L10)$~-$ 3.11($\pm$1.94) pc,
(c) log $SB($H$\alpha)$(this work) =  1.20($\pm$0.10) $\times~$
log $SB($H$\alpha)$(L10)$~-$ 0.11($\pm$0.07) \sb, 
and (d) \rat(this work) =  1.19($\pm$0.06) $\times~$\rat(L10)$+$ 3.07($\pm$0.89).

We estimated the diameters of the circles covering the region with \rat $>$ 0.4 
in emission line (\ratb) images.
In the cases of objects with partial shells,
we determined their sizes on the basis of the visible portion of the shells. 
The errors in measuring the sizes are about 0.5\arcsec~for small objects ($D$ = 10 $\sim$ 50 pc)
and about 1\arcsec~for larger objects (D $>$ 50 pc).
L10 determined elliptical regions that best traced optical shells of remnants, 
and they defined the geometric mean of the major and minor axes of above ellipses 
as sizes of the sources. They expanded slightly in a few cases to embrace X-ray emission 
that is apparently associated with the SNR.
The errors in their measuring the diameter are 0.5 $-$ 1\arcsec~for well-defined objects. 
In Figure \ref{comlo2}(b), we compared the diameters of the remnants in this work and L10. 
They agree well in general.  About 10 objects show differences larger than 10 pc.
We checked the images of these objects, and confirmed that our measurements are reasonable. 
The mean difference ($D$ = 10 $\sim$ 40 pc) is derived to be 1.7 pc, 
showing that our measurements are on average slightly larger than L10's.

\subsection{Classification Schemes for Progenitor Type and Morphological Type}

We applied the classification schemes for the progenitor type and 
the morphological type to the M33 SNR sample, as done for M31 in L14. 
The progenitor types are Type Ia SNRs and core-collapse (CC) SNRs.
Details are described in L14.
CC SNe are formed by high mass progenitors undergoing core collapse,
while Type Ia SNe are formed by white dwarfs in binary systems.
The former are typically located in star forming regions, 
while the latter can be found anywhere in a galaxy, regardless of their host galaxy types.
Thus, some previous studies examined the properties of stellar populations 
around SNRs to identify the nature of their SN progenitors \citep{chu88,bad09,fra12,jen12}. 
An existence of massive stars around an SNR candidate suggests that 
the remnant may result from a CC SN, 
while the absence of nearby massive stars suggests that it may be originated from a Type Ia SN.  
We selected the blue stars with $B-V<0.0$ and $V < 22.0$ mag ($\rm{M}_V < -2.5$ mag) 
from the photometric catalog given by \citet{mas06}.
Then we derived the number of OB stars in the surrounding region within 100 pc
for individual remnants, $N$(OB), following L14.
\citet{gva10} showed that some massive runaway stars 
have been detected at distances of about 100 pc 
from the center of star clusters in the Large Magellanic Cloud (LMC). 
Therefore the existence of OB stars within 100 pc from the center an SNR 
indicates that its progenitor might have been a massive star.
We adopted this distance as a boundary to count OB stars for progenitor classification.

\begin{figure}
   \plotone{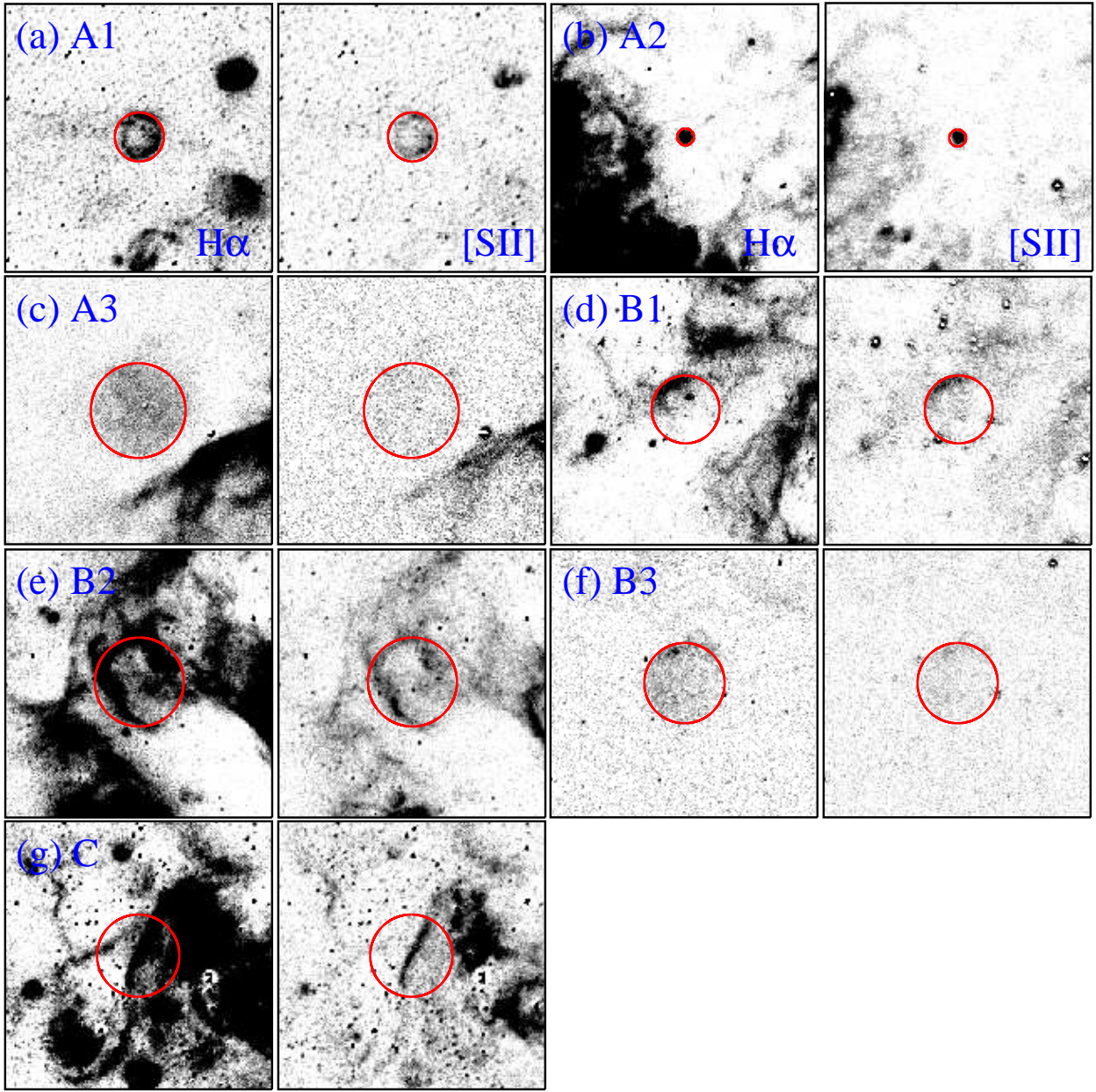}
   \caption{Samples of M33 remnants with different morphological categories.
            Sizes of the objects are marked by circles.
            An individual image covers $64\farcs8 \times 64\farcs8$ (251 pc $\times$ 251 pc). 
            North is up, and east to the left.}
\label{sample}
\end{figure}

\begin{figure}
   \epsscale{1}
   \plotone{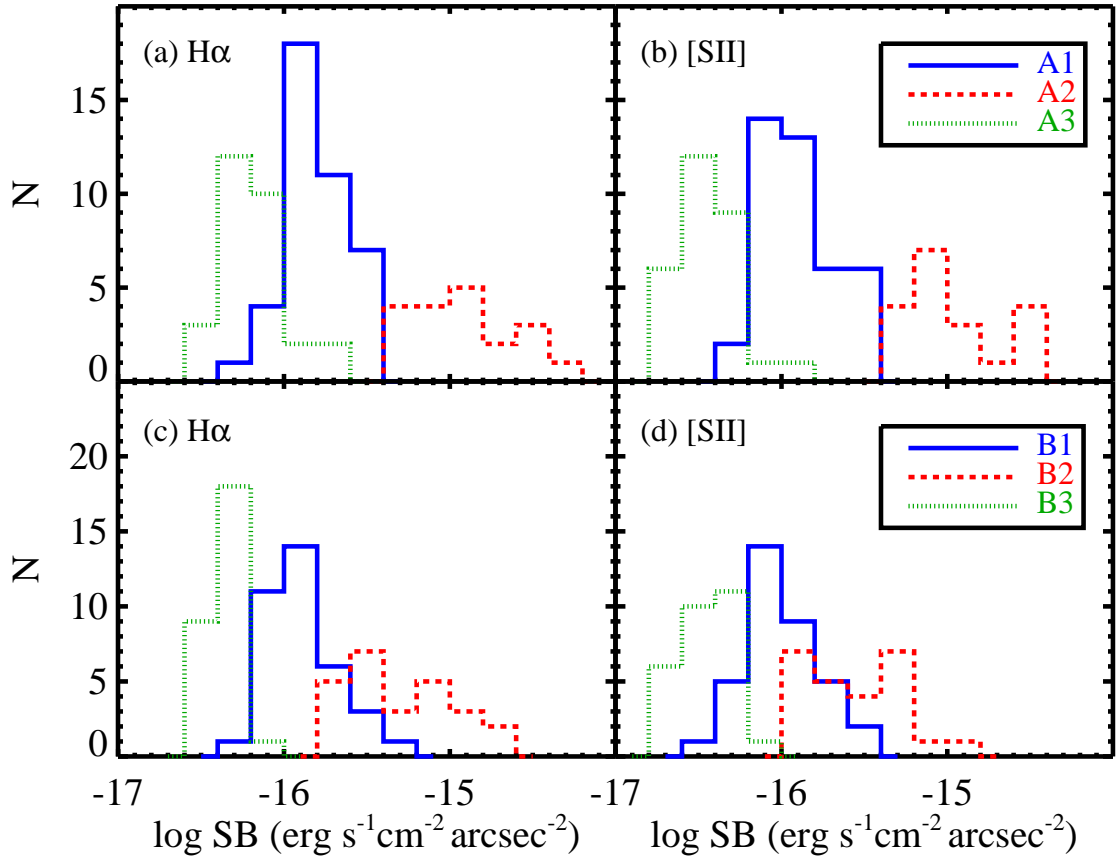}
\caption{Histograms of $SB$(\hab) (left panel) and $SB$(\s2) (right panel) 
          of M33 remnants with different morphological categories.}
\label{sb}
\end{figure}

Known SNRs have various morphological features:
a discrete shell, a filled-center nebula or a diffuse nebula. 
All the SNR samples in M33 show resolved morphological structure 
at the resolution of \ha images we use. 
Even in relatively confused regions on the outskirts of \h2
and SF regions the objects are resolved well.  
Thus we classified 199 sources in M33 
according to their shape as well as environments.   
Table \ref{table2} summarizes our criteria used for the classification.
Figure \ref{sample} displays some examples of the remnants
for individual morphological categories.
The majority of A2-class remnants have brighter $SB$ than other types,
as shown in Figures \ref{sb}(a) and (b).  
The majority of B2-class remnants are embedded in SF regions.
Their $SB$ is brighter compared with the other B-class objects, 
as displayed in Figures \ref{sb}(c) and (d). 

\section{RESULTS}

\subsection{A List of M33 Remnants}

By applying the SNR search method described in the previous section, 
we detected a total of 199 sources.
Among these there are 79 new objects detected in this work
and 120 known ones in the catalogs of previous studies (G98 and L10).
Table \ref{table3} presents a list of the 199 sources. 
In the table the columns give the ID number, the coordinates in right ascension
and declination (J2000.0), $L$, $D$, \ratb,  
morphological types, $N$(OB), and progenitor types.
The last column in Table \ref{table3} gives the identification matched with the catalog of L10. 

Of the sample, 170 objects result more likely from CC SNe, 
with the remainder to be from Type Ia SNe.
M33 has flocculent spiral arms so that a significant fraction of its disk 
is occupied by star forming regions. 
Therefore the Type Ia remnants located in the disk are likely to be classified 
as the CC remnants according to our classification criteria. 
Then the 170 sources classified as CC would be an overestimate.  
If the number density of the Type Ia remnants is uniform, about 10\% of the CC remnants
may be the Type Ia remnants. 
There are a few results for the relative ratio of two categories of remnants 
(Chu \& Kennicutt 1988; Franchetti et al. 2012; L14). 
According to the study of 32 objects in the LMC by \citet{chu88} 
more than $60\%$ of them are CC remnants. 
\citet{fra12} and L14 also inspected the stellar population around
remnants in M101 and M31, respectively.
They showed that $\sim$25\% (9 of 34) and $\sim$27\% (42 of 156) 
are more likely Type Ia remnants, respectively. 
A fraction ($\sim$17\%) of the Type Ia SNR remnants for M33 in this work 
is smaller than those for M101 and M31.  

The majority of the objects with A-class and B-class are CC SNRs ($\sim$85\% for each). 
In the cases of the A3-class ($\sim$31\%) and B3-class ($\sim$39\%) objects,
the Type Ia proportions are high, compared with those of objects of different morphological categories. 
L14 showed $\sim$85\% of M31 remnants are classified as shell remnants 
and $\sim$15\% are center-bright remnants. 
The majority ($\sim$90\%) of the M33 remnants are also shell remnants. 
It is noted that the proportions of the shell remnants and center-bright remnants in the MW 
at radio wavelengths are $78\%$ and $4\%$, while the rest is composite type remnants \citep{gre09}.

\subsection{Spatial Distribution}

\begin{figure}
   \epsscale{1.}
   \plotone{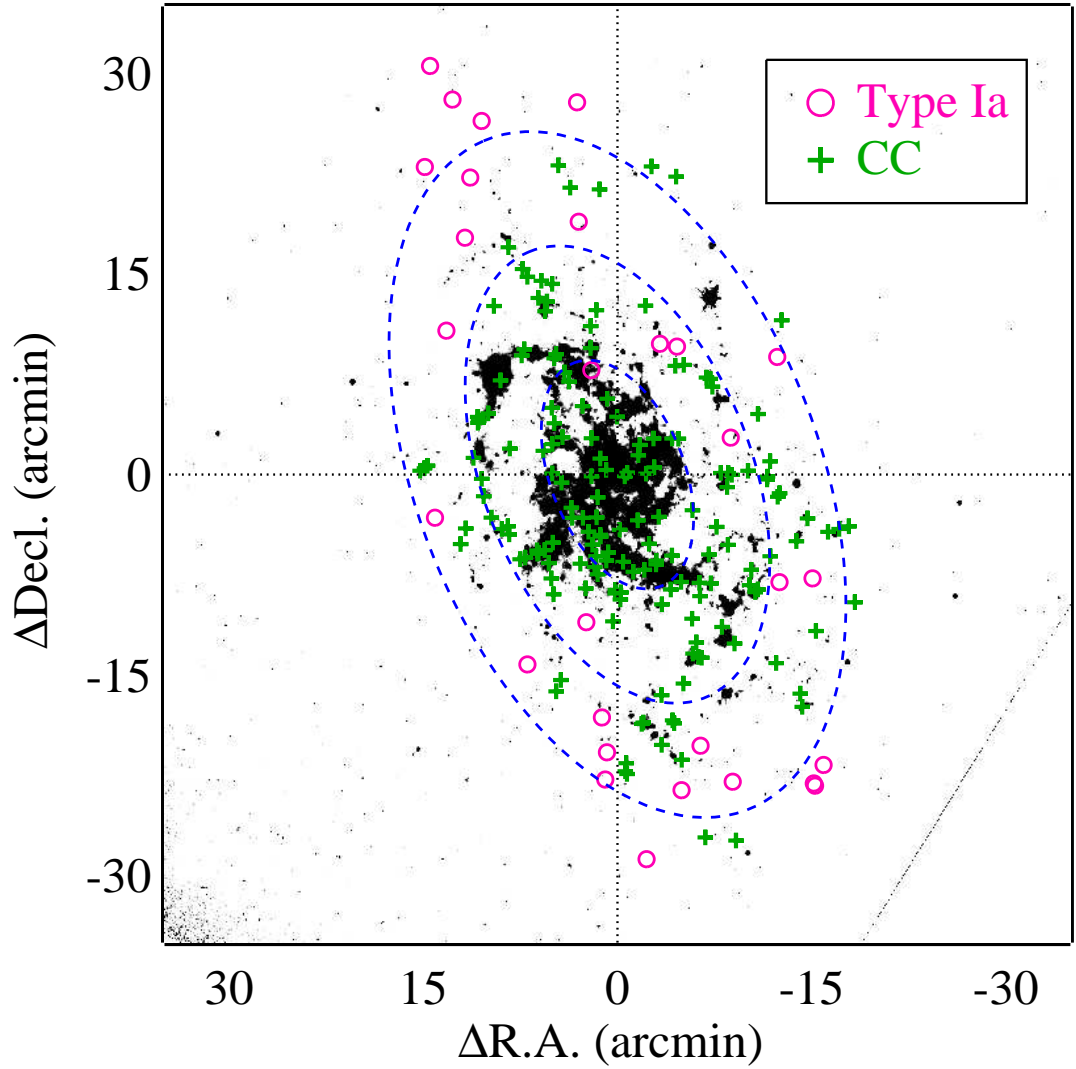}
\caption{Positions of Type Ia and CC remnants in M33
          plotted over the gray-scale map of the $Spitzer$ MIPS 24 $\mu$m image. 
          Dashed line ellipses represent 2, 4, and 6 kpc from the center.}
\label{spattype}
\end{figure}

Massive stars (which explode as CC SNe) would tend to be found near SF regions
such as \h2, molecular clouds, and spiral arms, 
while evolved, lower mass stars (which explode as Type Ia SNe) are 
expected to be distributed in a more random manner across the plane of a galaxy's disk. 
Below we examined the distributions of remnants relative to spiral arms as well as
the radial distributions of their number density. 
The positions of the M33 remnants are displayed in Figure \ref{spattype}.
The background image ($Sptizer$ 24 $\mu$m) exhibits the SF regions 
along the arms.
The CC remnants are mainly distributed along the spiral arms. 
In contrast the majority of the Type Ia remnants are located in the outer part of M33.

\begin{figure}
   \epsscale{1.}
   \plotone{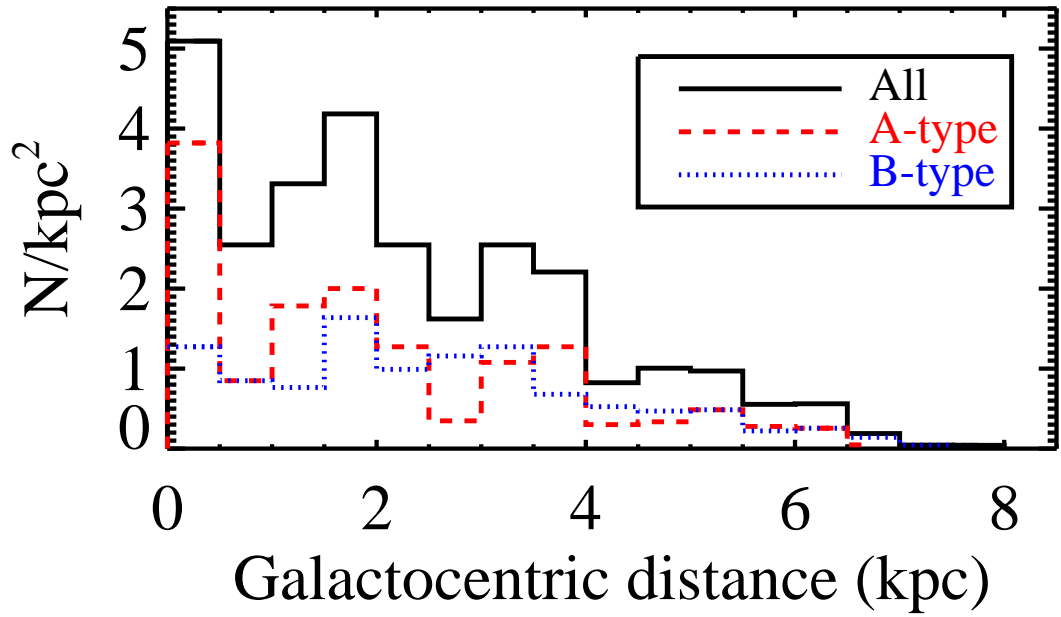}
\caption{Number density versus galactocentric distance of all, A-class, 
         and B-class remnants in M33.}
\label{spat2}
\end{figure}

The radial number density profile for the M33 remnants is seen in Figure \ref{spat2}. 
We derived deprojected galactocentric distance ($R$) 
adopting the position angle, $22.5^{\circ}$ \citep{pat03}
and the inclination angle, $56^{\circ}$ \citep{reg94}. 
The radial number density profile of all the objects decreases slightly outward 
with an abrupt break near $R$ = 4 kpc.  
Discernible concentrations are seen at $R \sim$ 2 kpc and $\sim$3.5 kpc, 
and a much weaker concentration at $R \sim$ 5 kpc.
The histogram of the A-class remnants shows discernible concentrations 
at the $R \sim$ 2 kpc and $\sim$3.5 kpc, and the number density decreases slightly outward. 
On the other hand, the distribution of the B-class remnants decreases more smoothly.
  
\begin{figure}
    \epsscale{0.7}
   \plotone{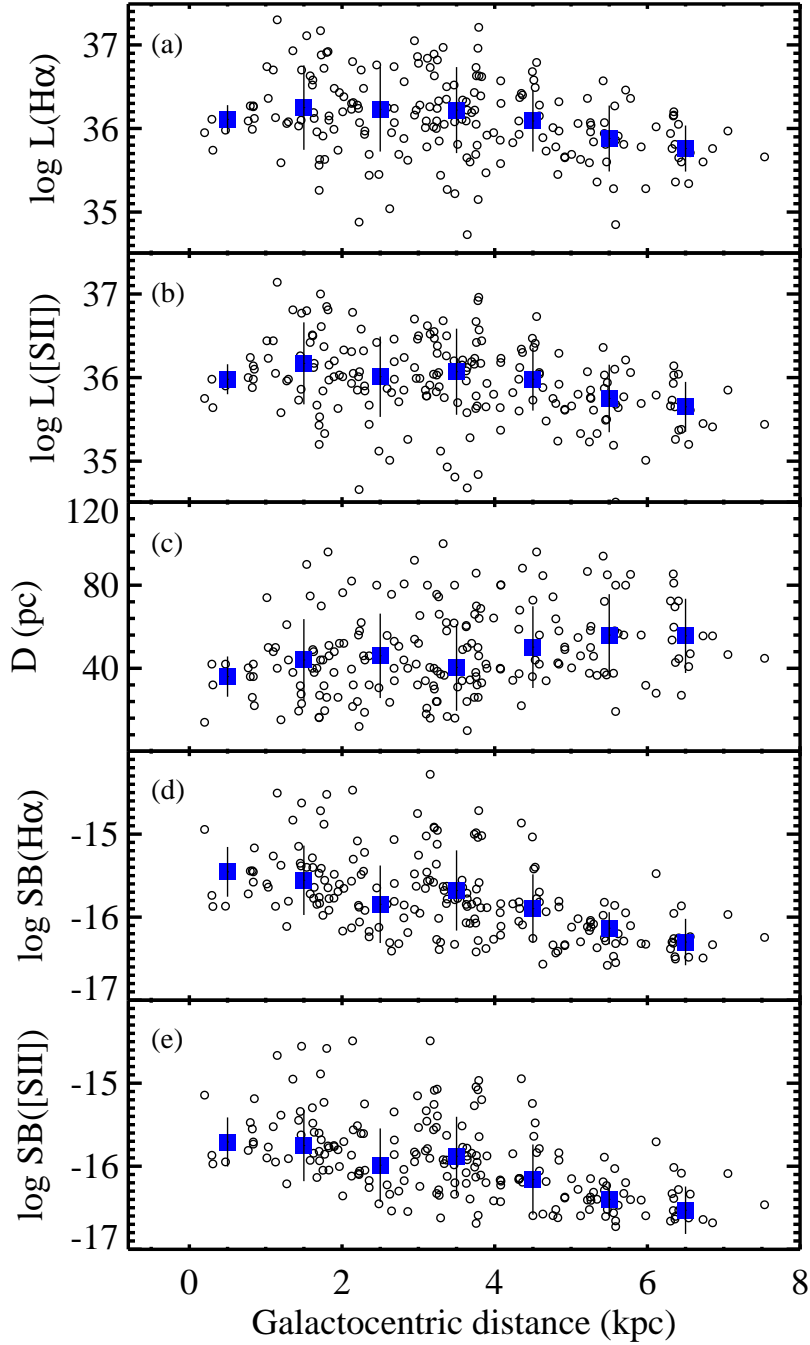}
   \caption{Radial distributions of (a) $L$(\hab) and (b) $L$(\s2), (c) $D$, 
            (d) $SB$(\hab) and (e) $SB$(\s2), of M33 remnants.
            Mean values are plotted by square symbols.}
\label{radialall}
\end{figure}

We investigated the radial variation in physical properties of all the remnants in 
Figure \ref{radialall}: (a) $L$(\hab) and (b) $L$(\s2), (c) $D$, (d) $SB$(\hab) and (e) $SB$(\s2).
The $L$(\hab) and $L$(\s2) of the objects change little at $R < 3.5$ kpc.  
In the outer region ($R > 3.5$ kpc),
$L$(\hab) and $L$(\s2) of the objects decrease slightly outward.
In the same range of $R$, $D$ increases from 40 pc to 50 pc.
Thus, the radial gradients of $SB$(\hab) and $SB$(\s2) are seen in Figures \ref{radialall} (d) and (e).     
This might be caused by the selection bias in our SNR survey.
We could find the remnants with more large $D$ and lower $SB$ easily
in the outer region of a galaxy. 

\subsection{Size Distribution} 

\begin{figure}
   \epsscale{0.8}
   \plotone{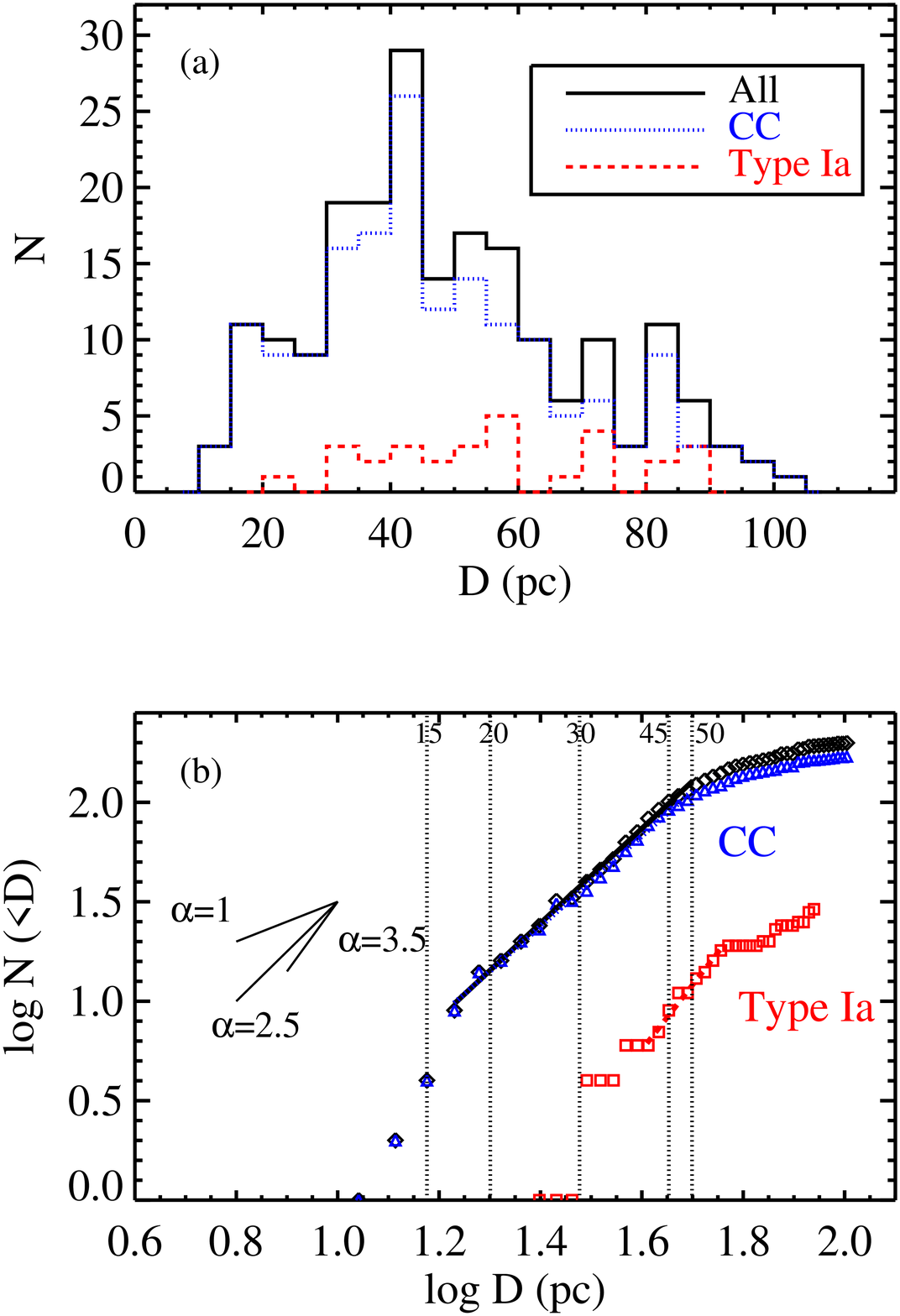}
\caption{(a) Differential and (b) cumulative size distributions of 
          all, CC, and Type Ia remnants in M33.}
\label{size}
\end{figure}

The evolution of SNRs has been a subject of many theoretical studies \citep{wol72,blo98,tru99}.
Some previous studies tried to explain observation data in theoretical aspects 
(Mathewson et al. 1983; Green 1984; Badenes et al. 2010, henceforth B10; Dopita et al. 2010).
Their evolution is described in terms of age. 
However, it is hard to determine observationally the age of the SNRs.  
Thus their sizes are used as a proxy for age. 
The differential size distributions of the M33 remnants are shown in Figure \ref{size}(a). 
Their diameters range from 10 pc to 100 pc. 
The histogram shows a noticeable concentration at $D \sim$ 45 pc with a broad wing toward the large size.
A discernible concentration at the same value is seen at the histogram of the CC remnants.
On the other hand, no distinguishable concentration is shown at the histogram of the Type Ia remnants. 
The majority of the objects smaller than $D =$ 30 pc are the CC remnants. 
The mean diameter of the Type Ia remnants is larger than that of the CC remnants. 
This means that a majority of the CC remnants may lie on denser ambient ISM than the Type Ia remnants.
                      
The $CSD$ of SNRs in a galaxy is useful for studying their evolutionary phases.                       
Their evolution is generally described by the three phase model:
the free expansion phase, the ST phase, and the radiative phase
\citep{wol72,mck77,dop03,dra11}. 
It is considered that the shapes of the $CSD$ are different, depending on the evolution phase.
If the SN rate in a galaxy is constant,  
the $CSD$ of SNRs becomes steeper as SNRs evolve.
During the free expansion phase, it will increase linearly, $N(<D) \propto D$. 
The power law index will be 2.5 for the ST phase,
and that will be 3.5 for the radiative phase.                                   
Thus, a study of the $CSD$ of SNRs in a galaxy 
is important to understand the evolution of SNRs. 
The transition sizes between these three phases 
are supposed to depend on several parameters including the density of surrounding ISM, 
mass of ejected material, and energy released by SNe \citep{blo98, tru99}. 
The transition radius from free expansion phase to ST phase
ranges from 1 to 10 pc, and that from ST to radiative phase is about 25 pc. 

The $CSD$ of the M33 remnants is seen in Figure \ref{size}(b). 
It follows a power law form ($N(<D) \propto D^\alpha$).
The power law index is $\alpha = 2.38 \pm 0.05$ ($D$ = 17 $\sim$ 50 pc). 
The majority of the remnants are considered to be in the ST phase. 
Their number decreases abruptly at $D =$ 17 pc 
as the size of the objects decreases.  
There are only a few sources that have sizes of $D = 10 \sim$ 17 pc. 
The slope of the $CSD$ of SNRs at $D >$ 50 pc 
is flatter compared with that of smaller SNRs. 
This indicates two possibilities.  
First, the flatter slope may be due to incompleteness.
We might have missed a significant number of the SNRs,
that have large $D$ and low $SB$ in the inner regions of the galaxy.  
Second, this may correspond to the transition stage from the ST phase to the radiative phase. 
B10 suggested that the majority of the objects are in the ST phase, 
and that they rapidly fade away below detection limit once they transit to the radiative phase.
The slope for the CC remnants is  $\alpha = 2.27 \pm 0.05$ ($D$ = 17 $\sim$ 50 pc),
but that for the Type Ia remnants is $\alpha = 3.26 \pm 0.12$ ($D$ = 40 $\sim$ 60 pc).

\begin{figure}
   \epsscale{0.8}
   \plotone{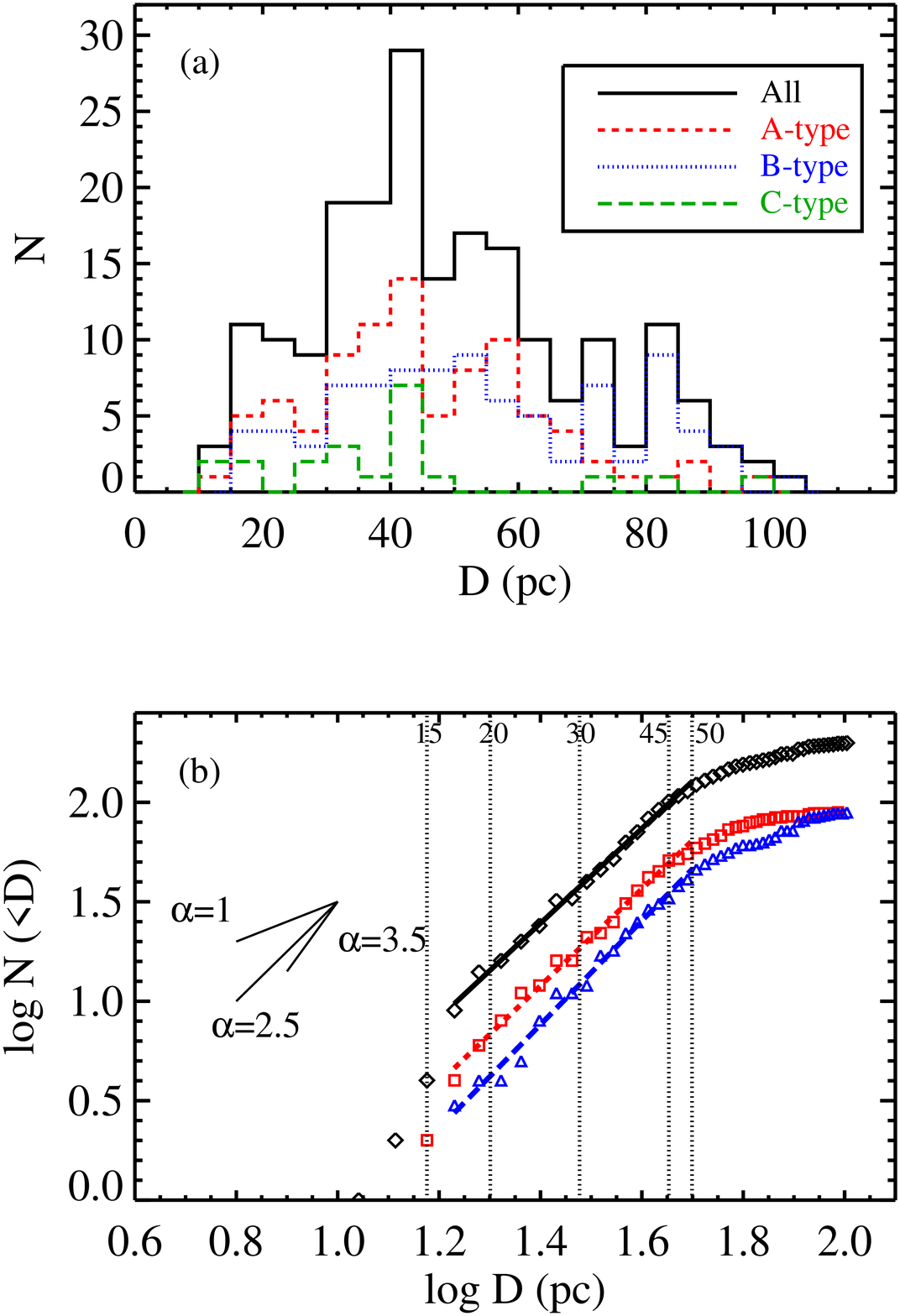}
\caption{(a) Differential and (b) cumulative size distributions of 
          all, A-class, B-class, and C-class remnants in M33.}       
\label{sizemor}
\end{figure}

In Figures \ref{sizemor}(a) and (b), we plot the size histograms of the objects 
with different morphological categories.   
The histogram of the A-class remnants shows two concentrations at $D \sim$ 40 pc and $\sim$60 pc,
whereas that of the B-class remnants 
shows a broad concentration at $D \sim$ 50 pc.
The diameter of the A-class remnants is, on average, $D \sim$ 45 pc, 
which is smaller than that of the B-class remnants, $D \sim$ 54 pc.
In Figure \ref{sizemor}(b), the $CSD$s of the A-class and B-class remnants
have close slopes up to 50 pc. 
The slopes for the A-class and B-class remnants
are $\alpha = 2.44 \pm 0.07$ ($D$ = 17 $\sim$ 50 pc) 
and $\alpha = 2.61 \pm 0.08$ for the same range, respectively. 
This means that the size evolution of SNRs may not depend on their morphological category. 
  
\subsection{\rat Distribution}

\begin{figure}
   \epsscale{0.9}
   \plotone{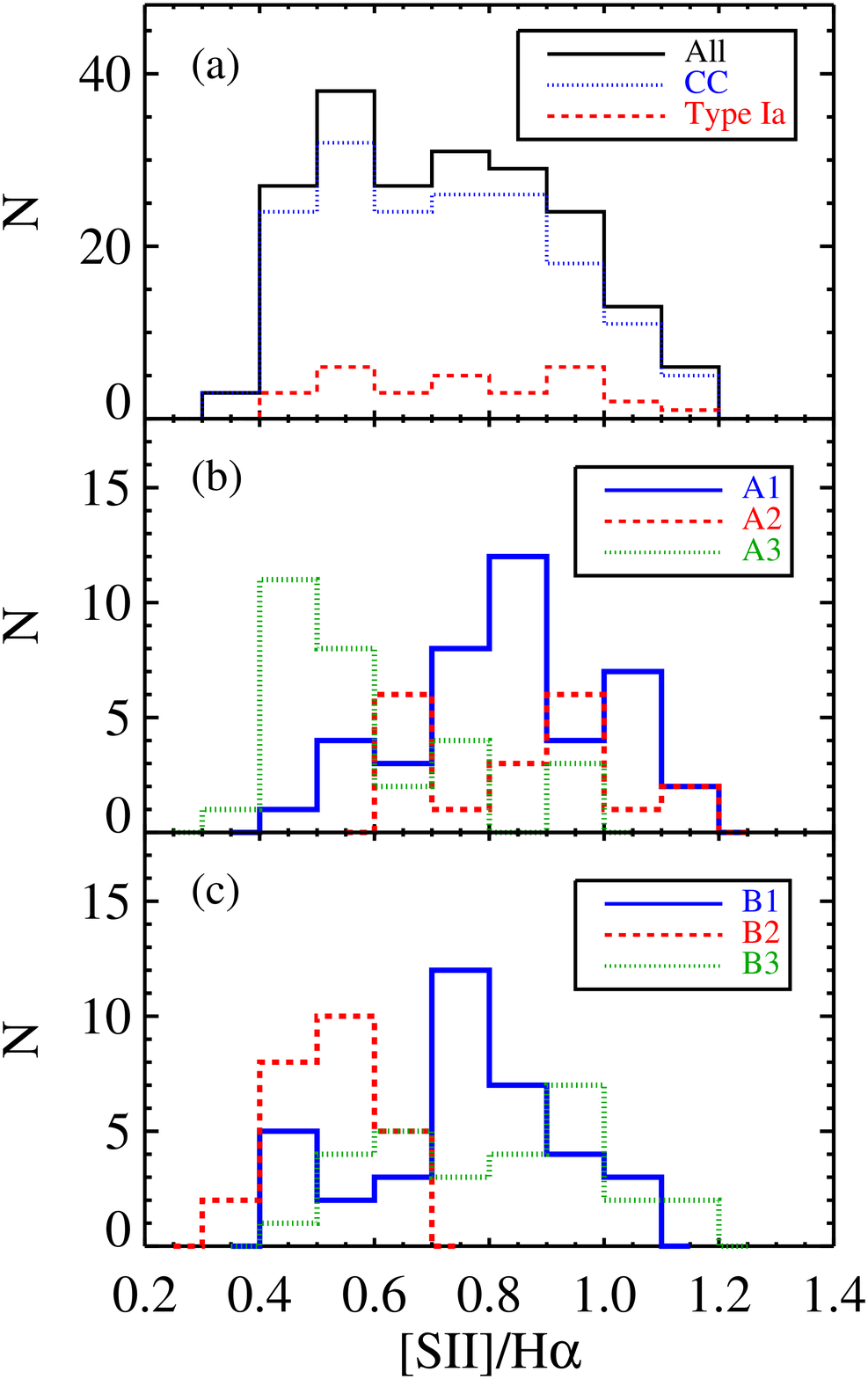}
\caption{Histograms of the \rat flux ratio of  M33 remnants:
        (a) different progenitor categories, (b) and (c) different morphological categories.}
\label{ratio}
\end{figure}

We derived the emission line ratio (\ratb) to identify the SNR candidates.
Objects with \rat $> 0.4$ were considered as SNRs in our survey. 
The histograms of emission line ratio (\ratb) for the sources 
with different progenitor categories are shown in Figure \ref{ratio}(a).
All the objects have \rat of 0.4 -- 1.2,
and their distribution is bimodal, with concentrations at \rat $\sim$ 0.55 and $\sim$0.8.
The histograms for different morphological categories in Figures \ref{ratio}(b) and (c)
are roughly divided into two sets at the boundary of \rat $\sim$ 0.7. 
Most of the A1-class, A2-class, B1-class, and B3-class remnants have elevated \ratb,
whereas most of the A3-class and B2-class remnants have low \ratb.
The A3-class remnants have low $SB$ and smooth shapes.
One may suggest that they should be diffuse ionized gas with large $D$ and low $SB$. 
Because their mean diameter is $D \sim$ 50 pc, 
we consider the objects to be bona-fide SNRs. 
The B2-class remnants with high $L$(\hab) and low \ratb are located in SF regions.

\subsection{$L - D$ Relation and $SB-D$ Relation}

\begin{figure}
   \epsscale{1.}
   \plotone{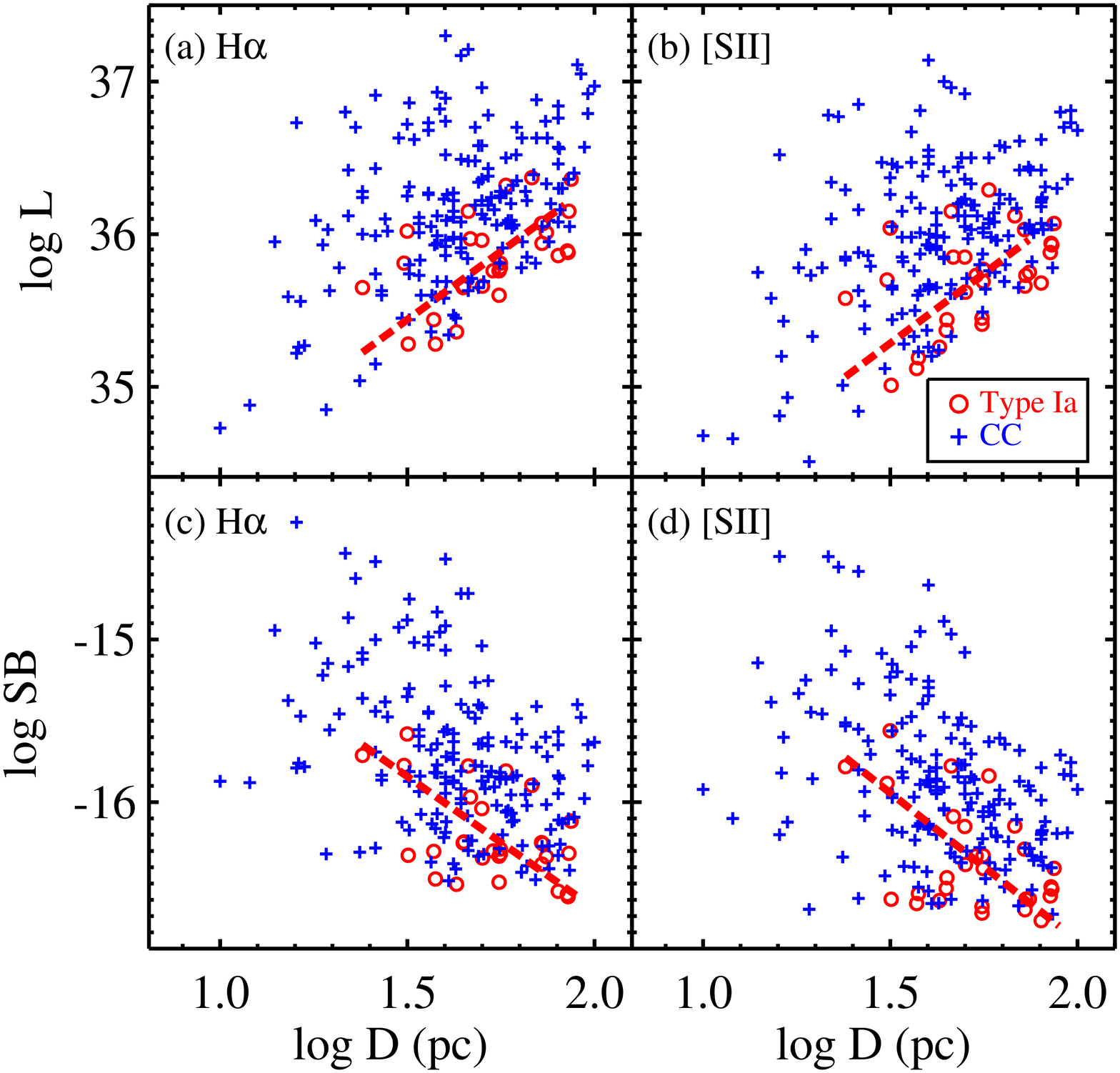}
   \caption{Upper panel: the relations between $L$ versus $D$ and
             lower panel : the relations between $SB$ versus $D$ 
             for Type Ia and CC remnants in M33.
             Dashed lines indicate linear least-squares fits.} 
\label{rel0}
\end{figure}

We examined the relations among $L$, $D$, and $SB$ of all, Type Ia, and CC remnants in M33. 
We inspected the relation between $L$ and $D$ of the remnants in Figures \ref{rel0}(a) and (b).
Weak correlations between $D$ and $L$(\hab) (and $L$(\s2)) of all the objects  
are found.
On the other hand, there are better correlations for the Type Ia remnants. 
We derived from fitting, $L\propto{D}^{1.78\pm0.29}$ 
and $L\propto{D}^{1.83\pm0.33}$ in \ha and \s2, respectively. 

The relation between $SB$ and $D$ for the Type Ia and CC remnants in M33 is seen 
in Figures \ref{rel0}(c) and (d). 
Weak correlations are found between $SB$(\hab) (and $SB$(\s2)) and $D$ of all the objects. 
The Type Ia remnants show better correlations 
between two parameters than the CC remnants.
For the Type Ia remnants, the slopes for $SB$ and $D$ relation 
in \ha and \s2 are $\beta = -1.62 \pm 0.39$ and $\beta = -1.81 \pm 0.53$, respectively.

\section{DISCUSSION}

\subsection{Comparison with Previous Studies of SNR Candidates in M33}

\begin{figure}
   \epsscale{1}
   \plotone{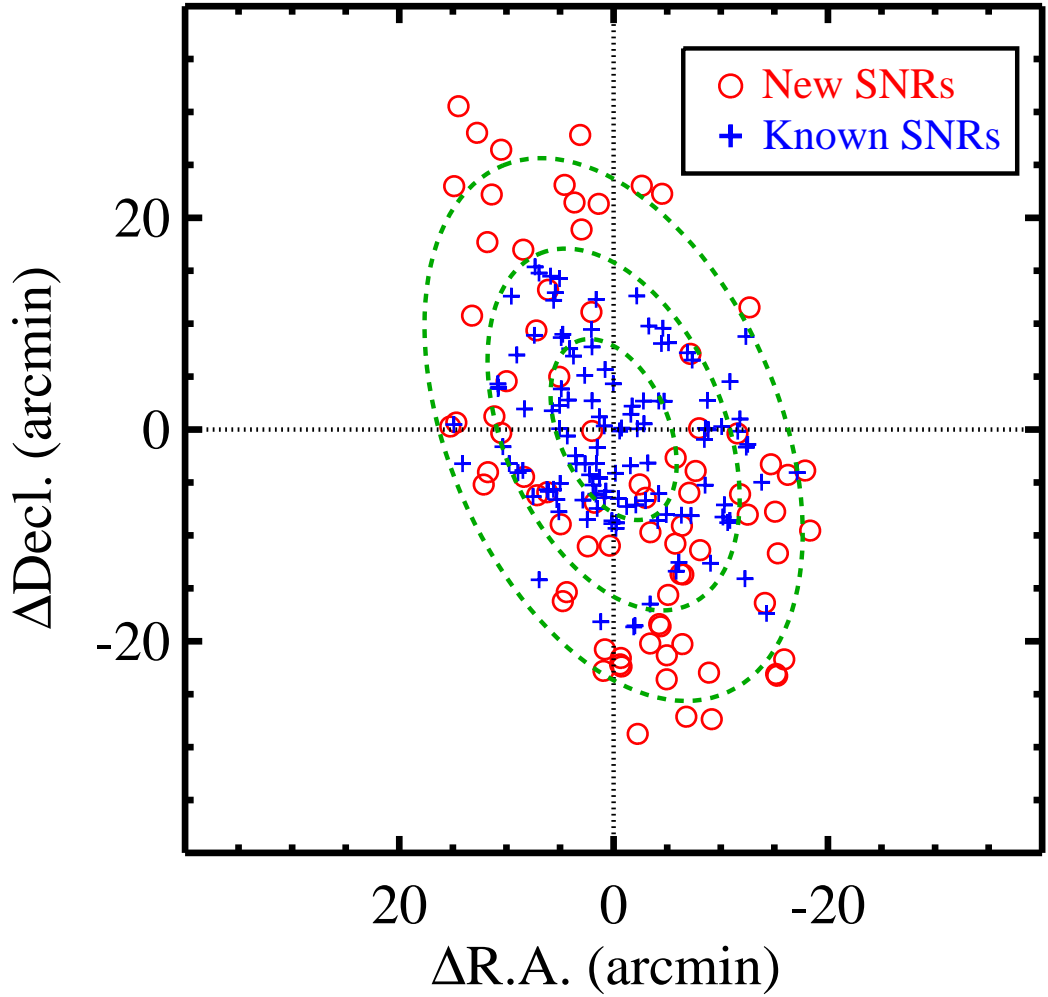}
\caption{Positions of newly found remnants and known ones in M33.
         Dashed line ellipses represent 2, 4, and 6 kpc from the center.}
\label{spatcom}
\end{figure}

There are several previous studies of M33 SNRs using optical images 
(Dodorico et al. 1978; Long et al. 1990; G98; L10).    
The number of known M33 SNRs detected in these studies is 137.
In this work we added 79 new ones, which have mostly low $SB$.  
We excluded 17 among the 137 known objects as mentioned in Section 2.1. 
These excluded objects are probably \h2 or superbubbles.
In Figure \ref{spatcom} we display the positions of the 79 new objects 
and the 120 known objects in M33. 
Previously known sources are mainly concentrated in the inner part of M33,
whereas the majority of the new ones are located in the outer part.
A majority of the new objects are considered to be relatively evolved sources
with low $SB$ at optical wavelengths, so that they are probably too faint to be detected in the previous X-ray surveys. 

\begin{figure}
   \epsscale{1.0}
   \plotone{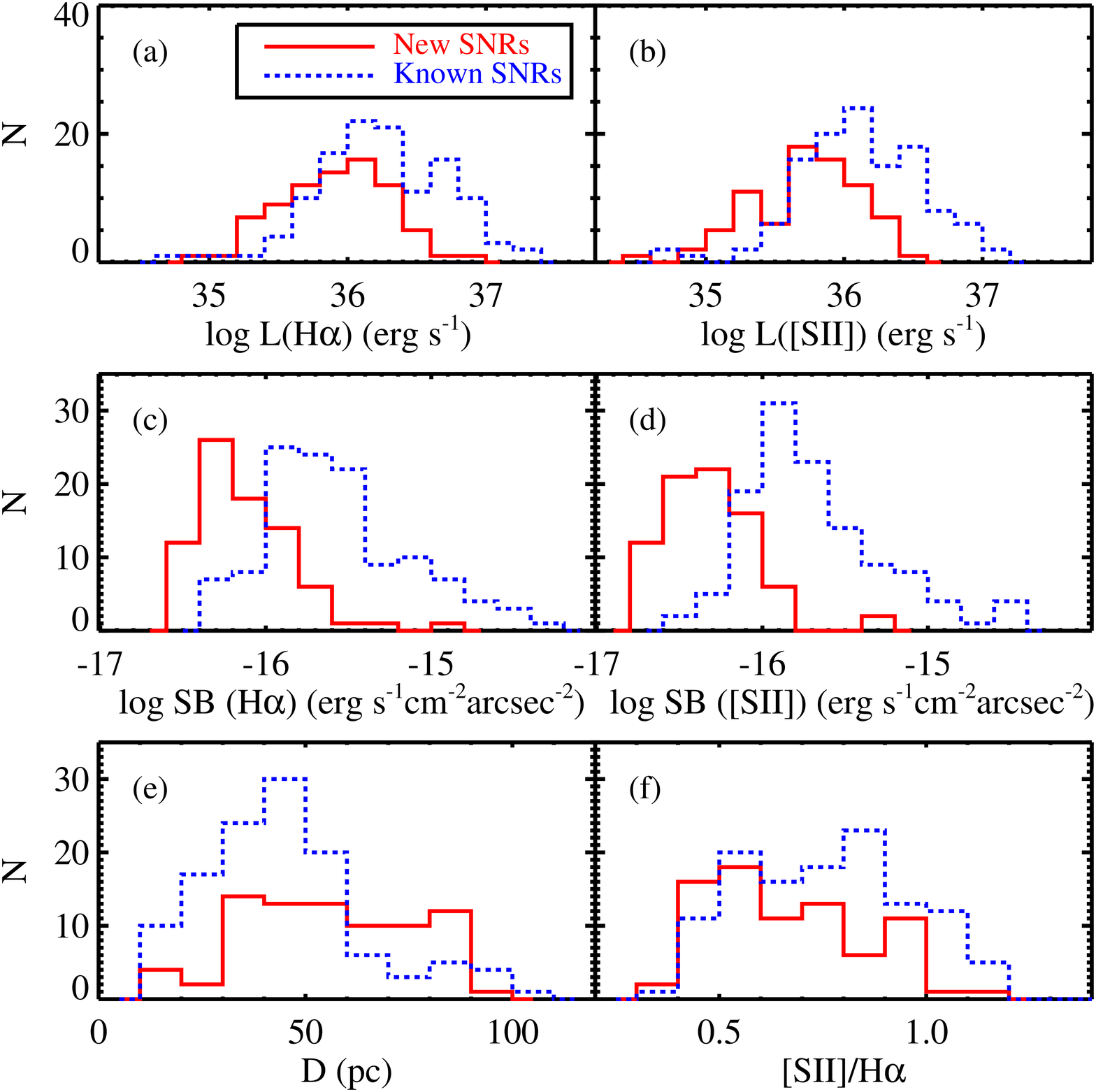}
\caption{Comparisons of histograms of (a) $L$(\hab) and (b) $L$(\s2),
         (c) $SB$(\hab) and (d) $SB$(\s2), (e) $D$, and (f) \rat of 
         new remnants with those of known ones in M33.}
\label{comlo}
\end{figure}

Figure \ref{comlo} displays the histograms of $L$(\hab), $L$(\s2), 
$SB$(\hab), $SB$(\s2), $D$, and \rat for the known SNR candidates and new ones in M33. 
The new objects have averagely fainter $L$, lower $SB$, and larger $D$ than the known ones. 
The known objects have \rat values ranging from 0.4 to 1.2,
with two concentrations at \rat $\sim$ 0.55 and $\sim$0.85.
The new ones have close \rat values, but with a noticeable concentration at \rat $\sim$ 0.5.    
The ratio of the number of the known objects to that of the new ones 
is lower for low \rat($<$ 0.7), than for elevated \rat($>$ 0.7).
A small number of the new objects with large $D$ are likely diffuse ionized gas. 
However, it is hard to distinguish bona-fide SNRs from diffuse ionized gas 
using only narrow band images.  

\begin{figure}
   \epsscale{1.}
   \plotone{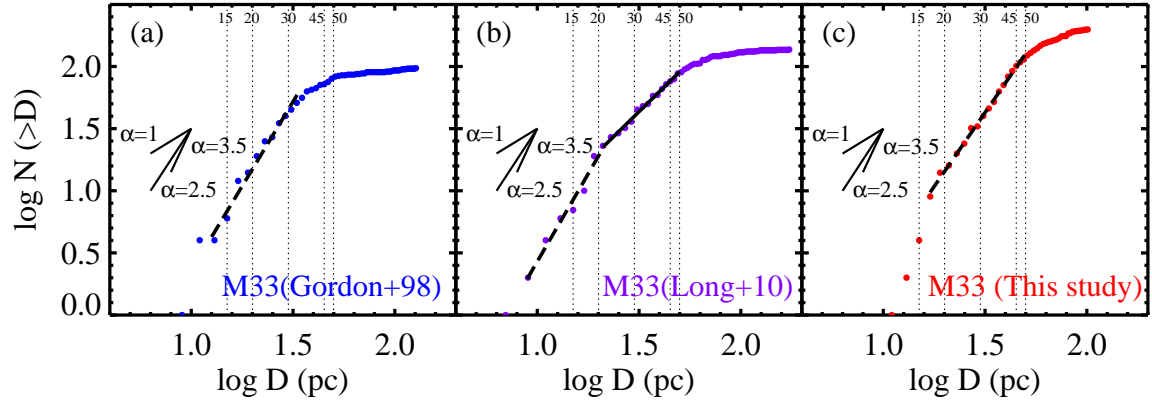} 
  \caption{Comparison of cumulative size distributions of remnants in this work 
         with those in the previous studies of M33: (a) G98, (b) L10, and (c) this work.
         Power law fits are represented by thick dashed lines.}         
\label{sdcom33}
\end{figure}

Figure \ref{sdcom33} compares the $CSD$ of the M33 remnants  
in this work with those in the previous studies (G98 and L10).
G98 presented the $CSD$ of 98 remnants
and showed that it follows a power law with $\alpha \sim$ 2.5.
They suggested that the majority of the objects are considered to be in the ST phase. 
In Figure \ref{sdcom33}(a), the slope of the $CSD$ obtained from the G98 sample 
is $\alpha = 2.72 \pm 0.14$ ($D$ = 13 $\sim$ 33 pc).  
The $CSD$ obtained from the L10 sample is shown in Figure \ref{sdcom33}(b). 
It follows a double power law with indices $\alpha = 2.82 \pm 0.04$ ($D$ = 9 $\sim$ 21 pc) and 
$\alpha = 1.60 \pm 0.03$  ($D$ = 21 $\sim$ 50 pc).
Figure \ref{sdcom33}(c) displays the $CSD$ of the remnants in this work.
It follows a power law, and the power index is $\alpha = 2.38 \pm 0.05$ ($D$ = 17 $\sim$ 50 pc).
If we adopt the same diameter ranges, $D$ = 19 $\sim$ 36 pc, for comparison, 
we obtain $\alpha = 2.23 \pm 0.08$ for this work, $\alpha = 2.24 \pm 0.13$ for the G98 sample, 
and $\alpha = 1.62 \pm 0.07$ for the L10 sample. 
Thus the value of the power index for our sample is very close to that of G98,
but is much larger than that of L10. 
The reason for the discrepancy in slope between this work and L10's 
is considered to be due to the difference in the size measurement.
Our measurement is on average slightly larger than L10's,
as described in Section 2.1.  
In conclusion, the majority of the M33 remnants are
considered to be in the ST phase, consistent with previous studies (Long et al. 1990; G98). 

\subsection{Comparison of SNR Candidates in M33 and M31}

The two spiral members of the Local Group galaxies, M31 and M33, 
have been the target of many observational studies to investigate the physical properties 
of these neighbouring systems. 
M31 and M33 are the nearest early-type (Sb) and late-type (Scd) spiral galaxies, respectively. 
Therefore, they may experience different star formation activities and have different ISM conditions. 
One might expect that there is a difference in the distributions of 
physical properties of SNRs in two galaxies. 

\begin{figure}
    \epsscale{1.0}
   \plotone{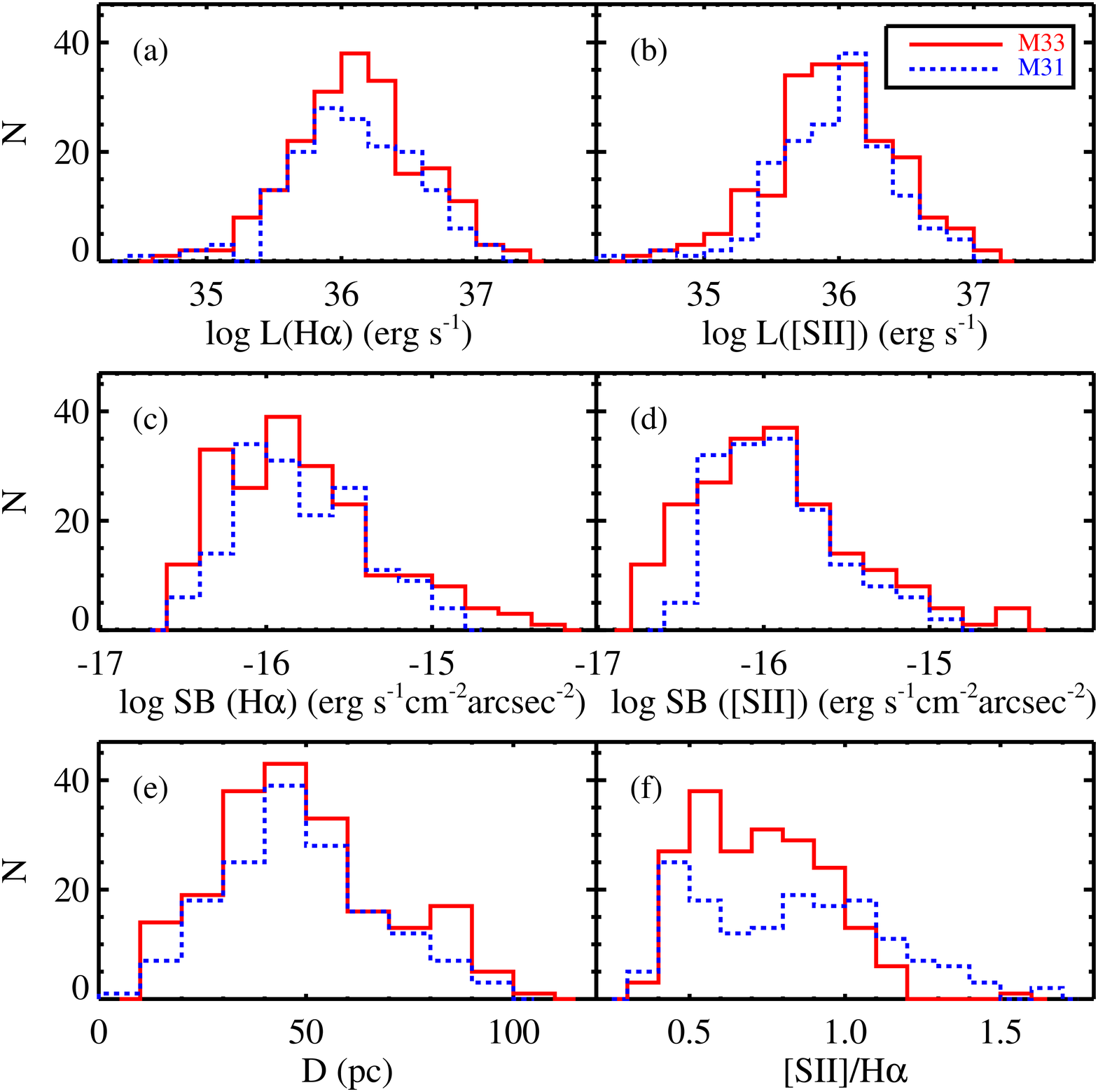}
   \caption{Comparisons of histograms of (a) $L$(\hab) and (b) $L$(\s2),
            (c) $SB$(\hab) and (d) $SB$(\s2), (e) $D$, and (f) \rat of M33 remnants 
            with those of M31 remnants (L14).}           
\label{compare2}
\end{figure}

We compared the physical properties of the M33 remnants in this work 
with those of the M31 remnants  studied using the similar methods by L14.
In Figures \ref{compare2}(a) and (b) we compare the histograms of $L$(\hab) and $L$(\s2), respectively.
The SNR detection limits for both M31 and M33 are found to be close.
The histograms of $SB$(\hab) and $SB$(\s2) have similar ranges as plotted in Figures \ref{compare2}(c) and (d).
Figure \ref{compare2}(e) compare the differential size distributions of the remnants in two galaxies. 
The histogram of the objects in M33 has a concentration at $D \sim$ 45 pc, close to that of the sources in M31. 
Figure \ref{compare2}(f) plots the histogram of emission line ratio (\ratb) of the remnants in M33 and M31. 
The histogram of the sources in M33 is bimodal, 
with two concentrations at \rat $\sim$ 0.55 and $\sim$0.8, 
and that of objects in M31 is also nearly bimodal,
having two concentrations at \rat $\sim$ 0.4 and $\sim$0.9.
In conclusion, there is little difference in the distributions of physical properties
between the M31 and M33 SNR candidates.

\subsection{Comparison of $CSD$s of the SNRs}

The $CSD$s of SNRs can be often described by the power law
\citep{mat83,mil84,gre84,hug84,lon90,dop10}.
The $CSD$s of both the MW SNRs and SNRs in the Magellanic Clouds (MCs)
were represented well by a power law index $\alpha \sim$ 1 \citep{mat83,mil84,hug84},
while the M33 remnants follow a power law, having an index $\alpha \sim$ 2.5 (G98).
L14 presented the $CSD$ of M31 remnants. 
It follows a power law, having an index $\alpha \sim$ 2.5.     
The $CSD$ is taken as evidence that the majority of the remnants 
in the MW and the MCs are likely to be in the free expansion phase,
while the majority of the objects in the M31 and M33 are considered to be in the ST phase. 

\begin{figure}
   \epsscale{1.}
   \plotone{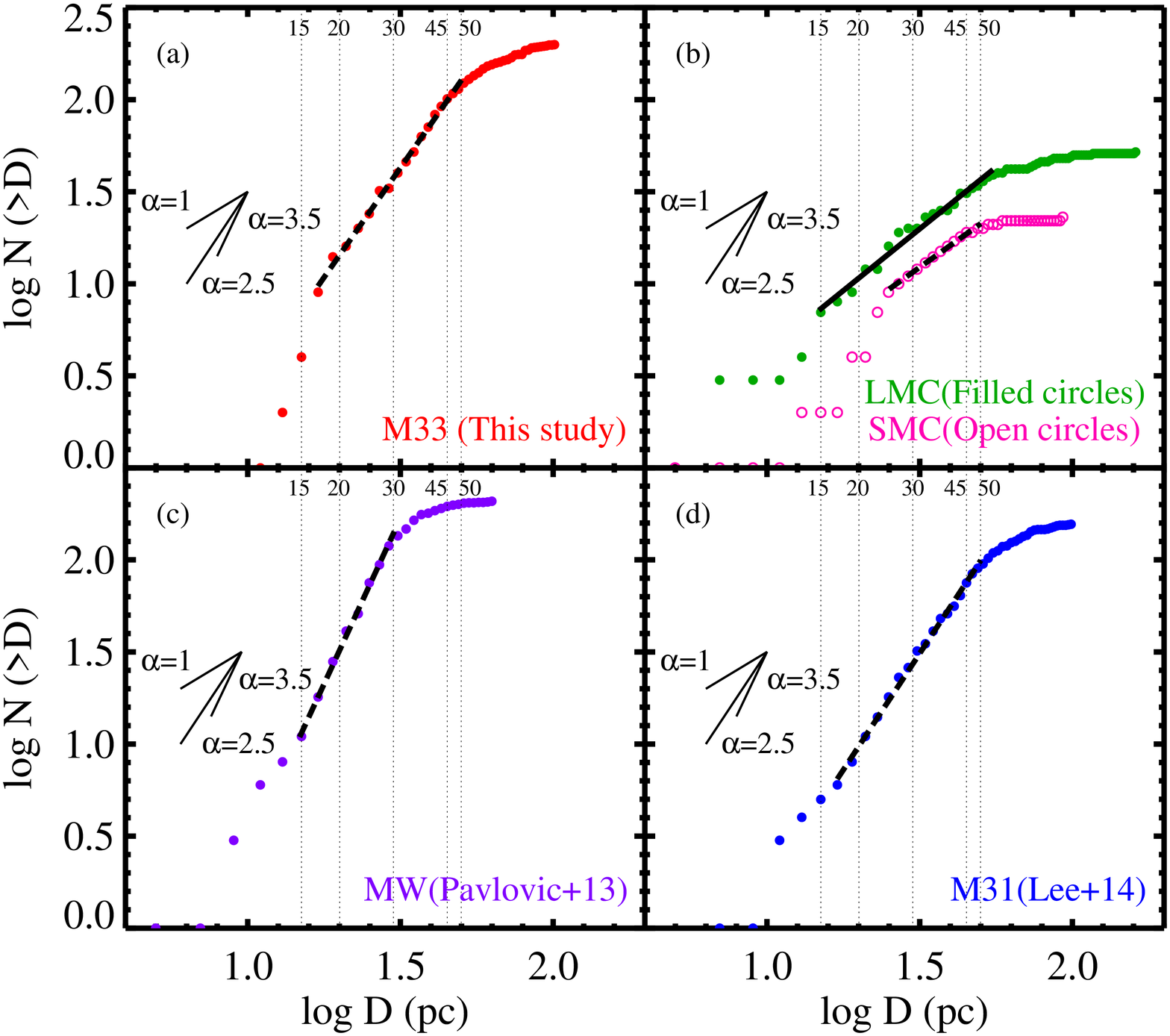} 
\caption{Cumulative size distributions of remnants in M33, the MCs, the MW, 
         and M31. Samples for the MCs, the MW, and the M31 
         are obtained from B10, \citet{pav13}, and L14, respectively. 
         Power law fits are represented by thick lines.
         The power law indices for individual galaxies are summarized in Table \ref{table4}.}
\label{sdcom}
\end{figure}

In Figure \ref{sdcom}, the $CSD$ of the M33 remnants
is compared with those for the MCs, the MW, and M31.
The $CSD$s for individual galaxies follow a power law. 
Table \ref{table4} presented the different power law indices among galaxies.
Figure \ref{sdcom}(a) displays the $CSD$ of the M33 remnants.
The power law index is $\alpha = 2.38 \pm 0.05$ ($D$ = 17 $\sim$ 50 pc). 
In Figure \ref{sdcom}(b), we display the $CSD$ of the remnants in the MCs.  
We derived the power law slopes using the B10 sample.
The value for the LMC SNRs is $\alpha = 1.34 \pm 0.04$ ($D$ = 15 $\sim$ 55 pc), 
and that for the SMC SNRs is $\alpha = 1.18 \pm 0.03$ ($D$ = 25 $\sim$ 50 pc). 
These results means that most of the MC SNRs are probably in the free expansion phase. 
However, it is noted that the shapes of the $CSD$s of SNRs depend also
on selection effect \citep{mil84} and the density distribution of ISM (B10). 

Figure \ref{sdcom}(c) plots the $CSD$ derived from the catalog of 274 MW SNRs \citep{pav13}. 
They estimated the sizes of most SNRs using the relation 
between radio $SB$ and $D$ derived from SNRs with known distances.
Thus the size estimates of these SNRs have large uncertainties.
The majority of the MW SNRs have diameters with $D = 15 - 30$ pc.
The slope for the size distribution is $\alpha = 3.60 \pm 0.06$ ($D$ = 15 $\sim$ 30 pc).
This is much steeper than the value for M33, 
indicating that most of the MW SNRs are likely to be in the radiative phase.
The $CSD$ of the objects in M31 obtained from L14 is seen in Figure \ref{sdcom}(d). 
They presents the diameters of the 156 sources in M31 detected in the LGGS data. 
The slope of the $CSD$ of these SNRs is derived to be 
$\alpha = 2.53 \pm 0.04$ ($D$= 17 $\sim$ 50 pc). 
This value is very close to that for the objects in M33.

\subsection{Multiwavelength Properties of Remnants}

Inspecting the properties of SNRs in multiwavelength images 
is useful for understanding their nature and evolution.
Some previous studies for extragalactic SNRs showed weak or little correlations 
among X-ray, radio, and optical properties of SNRs (Pannuti et al. 2007; L10; Leonidaki et al. 2013). 
On the other hand, a good correlation 
between $L_{\rm X}$ and $L$(\hab) for the center-bright remnants in M31
were found by L14. 
Here we discuss any correlations between X-ray (radio) and optical properties of the M33 remnants.

\begin{figure}
   \epsscale{0.8}
     \plotone{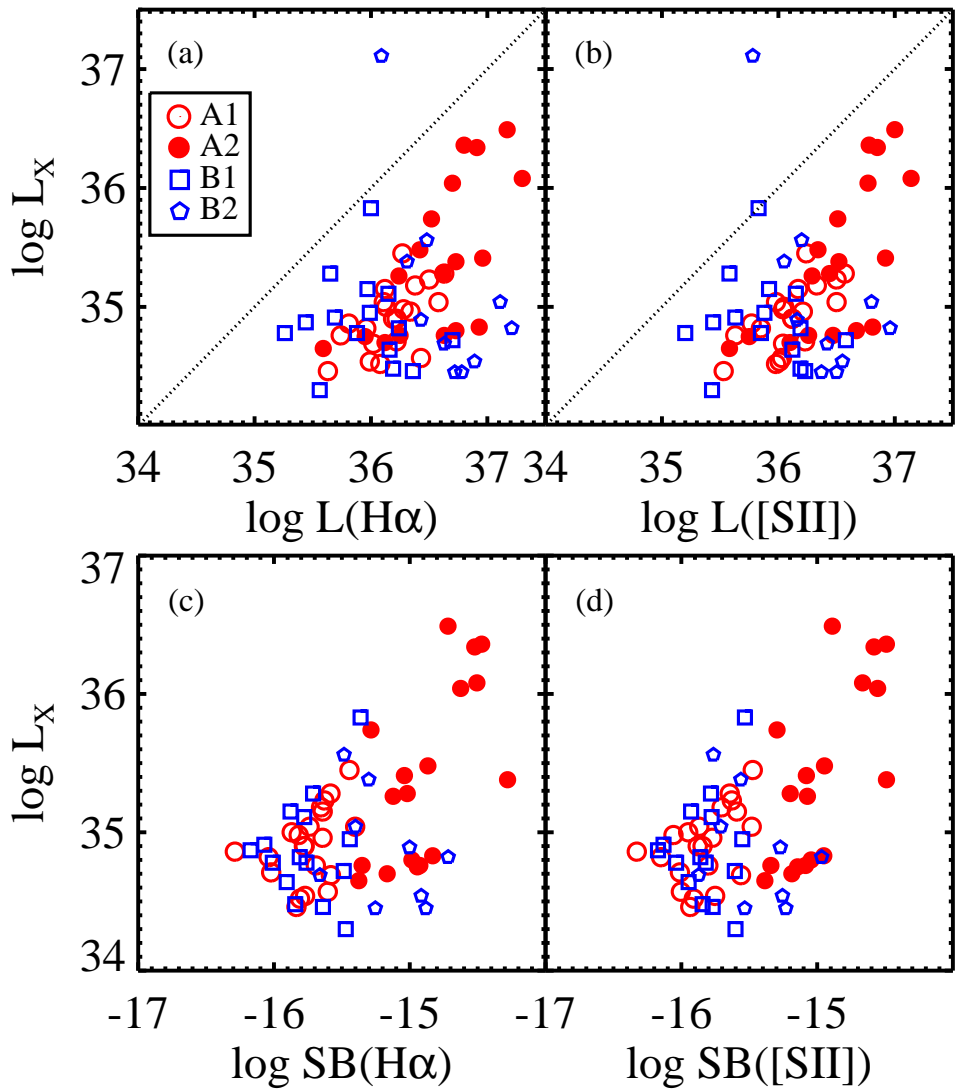} 
\caption{Comparisons of $L_{\rm X}$ and optical properties of remnants in this work and L10 for M33:
         (a) $L_{\rm X}$ versus $L$(\hab), (b) $L_{\rm X}$ versus $L$(\s2),
         (c) $L_{\rm X}$ versus $SB$(\hab), and (d) $L_{\rm X}$ versus $SB$(\s2).}
\label{multi1}
\end{figure}

First, the optical properties of the M33 remnants
are compared with their property at X-ray wavelengths. 
L10 detected 82 X-ray bright sources among the 137 optically selected samples 
using $Chandra$ data from the ChASeM33 survey. 
Of 199 sources detected in this work, 78 are matched with L10 catalog. 
Much larger proportions of the remnants with complete shells 
($\sim$51\% and $\sim$95\% for A1-clsss and A2-class remnants, respectively) 
are found at the X-ray wavelengths. 
In Figures \ref{multi1}(a) and (b), $L$(\hab) and $L$(\s2) of the matched sources
are compared with $L_{\rm X}$, respectively. 
All the matched sources with an exception 
are brighter at optical wavelengths than at X-ray wavelengths. 
Weak correlations between the optical luminosity ($L_{\rm opt}$) and $L_{\rm X}$ are found
for all the matched objects.
In the cases of the A2-class remnants, 
we calculated the linear Pearson correlation coefficients 
in \ha and \s2, obtaining 0.66 and 0.69 
(corresponding probabilities for given numbers of the sample are $\sim$99\% for each). 
They show that the two sets of quantities have strong correlations.

In contrast, there are little correlation between $L_{\rm opt}$ and $L_{\rm X}$ of the B-class SNRs. 
The correlation difference between the A-class remnants and 
the B-class remnants is considered to be related with the distribution of the ambient ISM. 
\citet{pan07} pointed out, based on simple emissivity models, 
that $L_{\rm opt}$ and $L_{\rm X}$ of SNRs might 
increase with increasing density of the ambient ISM. 
They expected a good correlation between $L_{\rm opt}$ and $L_{\rm X}$ of SNRs,
if the ambient ISM is uniform.
The correlation between $L_{\rm opt}$ and $L_{\rm X}$ of the A-class remnants 
is stronger than that of the B-class remnants in M33. 
This means that the ISM around the A-class remnants
is more uniform than that around the B-class remnants. 

Figures \ref{multi1}(c) and (d) show $L_{\rm X}$ versus $SB$(\hab) and $SB$(\s2),
respectively, of the matched sources.  
There are tight correlations between $SB$ of only the A2-class remnants and $L_{\rm X}$. 
Correlation coefficients are derived to be 0.63 (and 0.69) in \ha (and \s2). 
Thus the two samples are considered to be strongly correlated.

\begin{figure}/
   \epsscale{0.8}
     \plotone{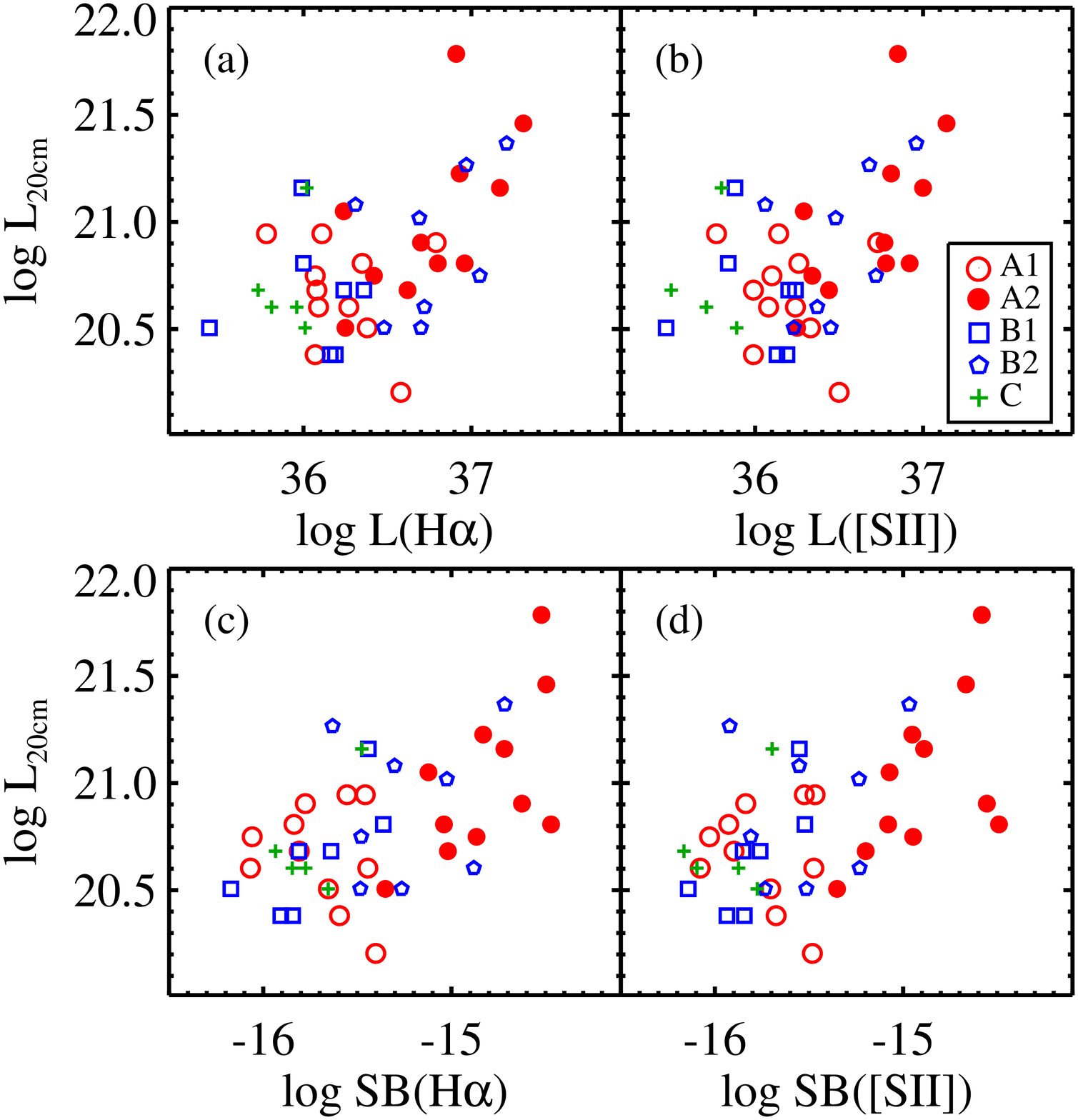} 
\caption{Comparisons of $L_{\rm 20cm}$ and optical properties of remnants in this work and \citet{gor99} for M33:
         (a) $L_{\rm 20cm}$ versus $L$(\hab), (b) $L_{\rm 20cm}$ versus $L$(\s2),
         (c) $L_{\rm 20cm}$ versus $SB$(\hab), and (d) $L_{\rm 20cm}$ versus $SB$(\s2).}
\label{multi2}
\end{figure}

Second, we inspected the correlation between optical and radio properties of the M33 remnants. 
\citet{gor99} presented a catalog of 186 radio sources obtained from $VLA-WRST$ observations. 
We found that 43 of the 199 optically selected samples in this work
are matched with \citet{gor99} catalog. 
Higher fractions of the A2-class ($\sim$63\%) and B2-class ($\sim$42\%) remnants 
are detected than other categories at radio wavelengths as well as at optical wavelengths. 
Figures \ref{multi2}(a) and (b) show comparisons of $L$(\hab) and $L$(\s2) 
of the matched sources with $L_{\rm 20cm}$, respectively.
Weak correlations between $L_{\rm opt}$ and $L_{\rm 20cm}$
for all the matched objects are found,
but strong correlations for the A2-class remnants. 
In Figures \ref{multi2}(c) and (d), we plot $L_{\rm 20cm}$ versus
$SB$(\hab) and $SB$(\s2), respectively, of the matched sources. 
We found that there are also good correlations 
between $SB$(\hab) and $L_{\rm 20cm}$.

\begin{figure}
   \epsscale{0.9}
   \plotone{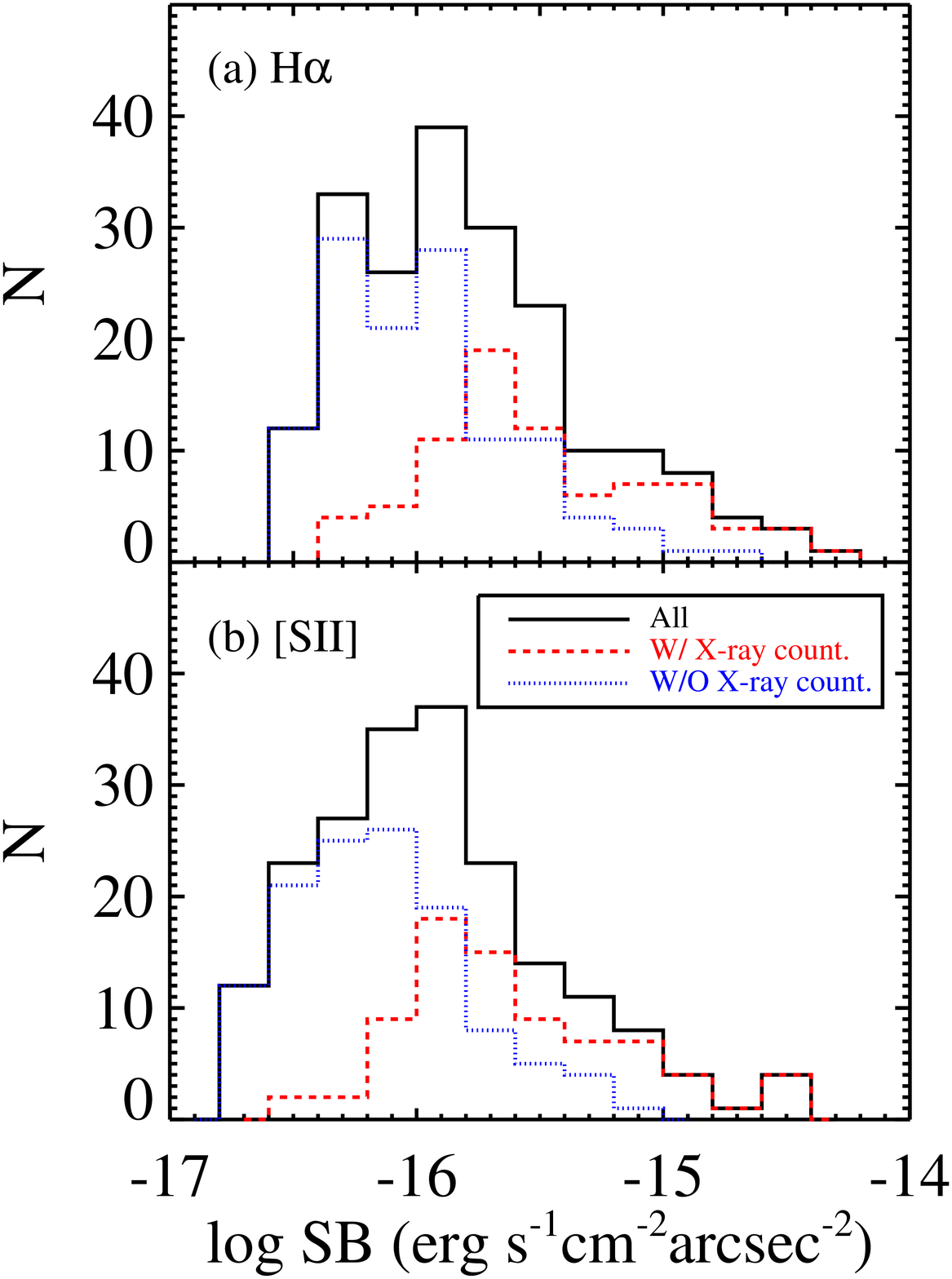}
\caption{Histograms of (a) $SB$(\hab) and (b) $SB$(\s2) of remnants with X-ray emission,  
         those without X-ray emission, and all the M33 remnants.}       
\label{sbx}
\end{figure}

\begin{figure}
   \epsscale{0.8}
   \plotone{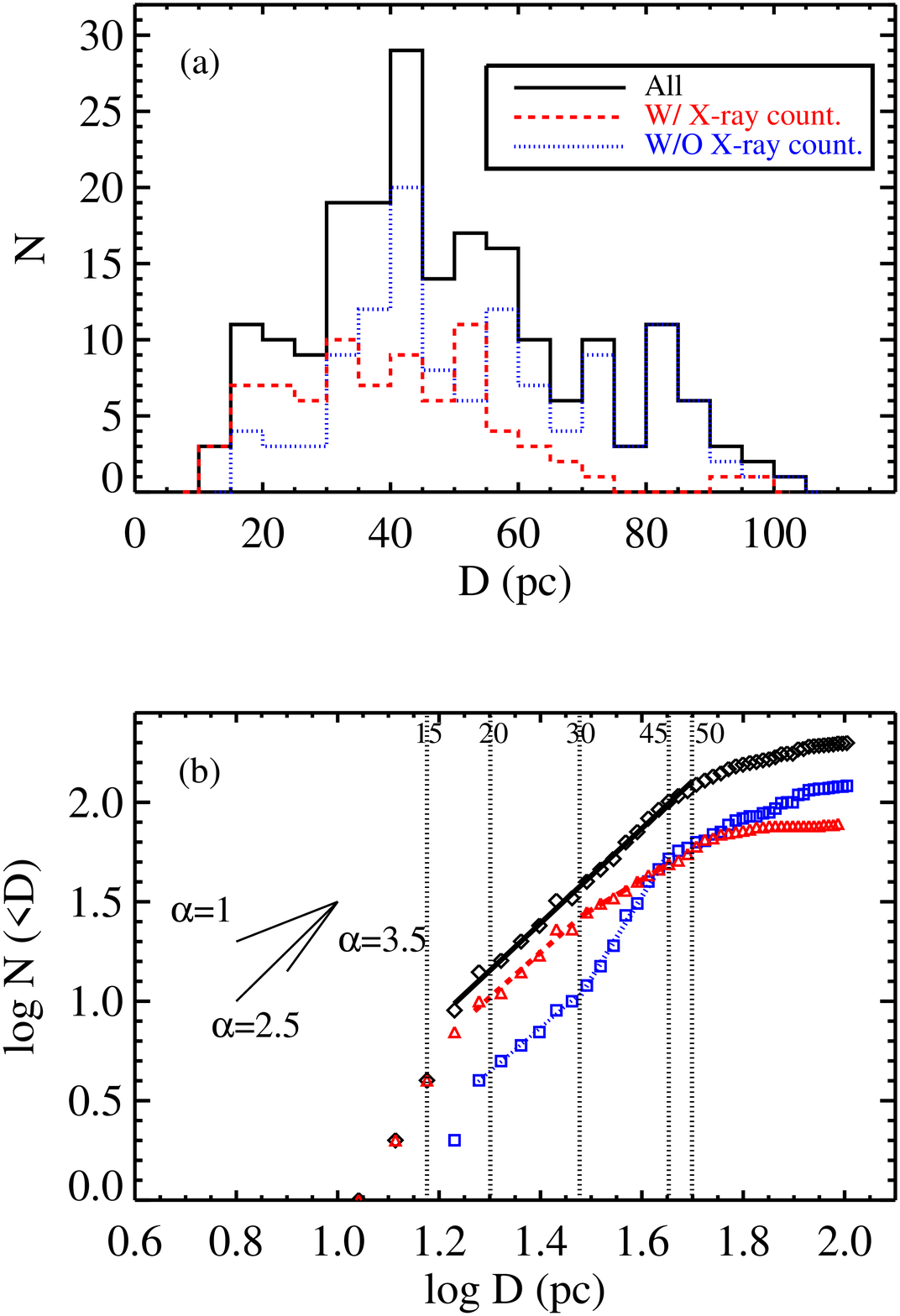}
\caption{(a) Differential and (b) cumulative size distributions of remnants 
          with X-ray emission, those without X-ray emission, and all the M33 remnants.}        
\label{sizex}
\end{figure}

The histograms of $SB$(\hab) and $SB$(\s2) for the objects with X-ray emission, those without X-ray emission, and all the M33 remnants are seen in Figure \ref{sbx}.  
The majority of the objects with large $SB$(\hab) (and $SB$(\s2)) have X-ray counterparts.
The distributions in \ha and \s2 of the objects with X-ray emission 
show concentrations at $10^{-15.7}$ \sb~ and $10^{-15.9}$ \sb, respectively. 
The majority of the objects with low $SB$ have no X-ray counterparts.
The objects with X-ray emission have a nearly flat distribution for $D = 15 - 60$ pc 
as shown in Figure \ref{sizex}(a).
The objects with X-ray emission have a median $D$ of 38 pc,
smaller than the value of 50 pc for those without X-ray emission. 
Figure \ref{sizex}(b) shows that the slope for the objects with X-ray emission
is $\alpha = 2.26 \pm 0.2$ ($D$ = 19 $\sim$ 30 pc), 
while that for those without X-ray emission is $\alpha = 2.21 \pm 0.05$.
However, the slopes of the $CSD$ of the remnants for 30 $ < D <$  46 pc are different. 
The slopes for the objects with X-ray emission is $\alpha = 1.52 \pm 0.02$,
much flatter than those without X-ray emission, $\alpha = 4.11 \pm 0.11$.

\section{SUMMARY} 

We presented a survey of SNR candidates in M33 using the \rat technique 
to the optical narrow band images.
We expanded the remnant list to 199 objects, of which 79 are new findings. 
We utilized this catalog to study their optical and X-ray properties.
We found that the majority of them are the remnants of CC SNe.
Type Ia proportion ($\sim$15\%) for M33 is lower than that for M31 ($\sim$27\%) (L14).  
The radial number density profile of the remnants
shows two significant concentrations at $R \sim$ 2 kpc and $\sim$ 3.5 kpc. 
For $R >$ 3.5 kpc, the sizes of the objects increase, on average, slightly outward.
In contrast the mean value of their $SB$(\hab) (and $SB$(\s2)) decreases. 
A noticeable concentration is seen at $D \sim$ 45 pc 
in the differential size distribution of the CC remnants.
Power law fitting to the $CSD$ of the remnants yielded a value,
$\alpha = 2.38 \pm 0.05$ (for the range of $D=17 -- 50$ pc).
This value is close to the value from that for the M31 remnants, $\alpha = 2.53 \pm 0.04$ (L14). 
Thus the majority of the objects in both M33 and M31 are the ST phase. 
The histogram of the emission line ratio (\ratb) of the remnants
has two concentrations, at \rat $\sim$ 0.55 and $\sim$0.8.
Tight correlations between $D$ and $L$(\hab) (and $L$(\s2))
for the Type Ia remnants are found. 
A significant proportion of the remnants
are detected in X-rays for the A1-class and A2-class ($\sim$51\% and $\sim$95\%, respectively).  
These sources have mostly large $SB$, small $D$, and complete shapes at optical wavelengths. 
Strong correlations between $L_{\rm X}$(and $L_{\rm 20cm}$) 
and optical properties are found for the A2-class remnants.

This work was supported by a National Research Foundation of Korea (NRF)
grant funded by the Korea Government (MSIP) (No. 2012R1A4A1028713).

\clearpage

\begin{deluxetable}{cccccccc}
\tabletypesize{\small}
\tablecaption{A List of Previous M33 SNR Candidates Excluded in This Work  \label{table1}}
\tablewidth{0pt}
\tablehead{
\colhead{Name} & \colhead{\footnotesize{R.A.(J2000.0)\tablenotemark{a}}} &
\colhead{\footnotesize{Decl.(J2000.0)\tablenotemark{a}}} &
\colhead{\footnotesize{log $L$(\hab)\tablenotemark{b}}}    &
\colhead{\footnotesize{log $L$(\s2)\tablenotemark{b}}}     &
\colhead{$D$\tablenotemark{c}}   &   \colhead{\footnotesize{\ratb}}  &
\colhead{\footnotesize{Class.}\tablenotemark{d}}   \\
  & \colhead{\footnotesize{(Degree)}}    & \colhead{\footnotesize{(Degree)}}        & 
\colhead{\footnotesize{(erg $s^{-1}$)}}  & \colhead{\footnotesize{(erg $s^{-1}$)}}  &
\colhead{(pc)} &  }
\startdata
L10-1 &   23.1262779 &     30.462805 & 37.02 & 37.01 & 136 &   0.98 &    S \\
L10-3 &   23.1772499 &     30.349541 & 36.77 & 36.58 &  98 &   0.64 &    S \\
L10-12 &   23.2502728 &     30.512455 & 37.49 & 37.07 &  62 &   0.38 &    H \\
L10-19 &   23.2814598 &     30.714882 & 36.66 & 36.21 &  74 &   0.35 &    H \\
L10-28 &   23.3278923 &     30.451220 & 37.10 & 36.84 & 158 &   0.55 &    S \\
L10-35\tablenotemark{e} &   23.3708439 &     30.795719 & 35.52 & 34.65 &  18 &   0.13 &      \\
L10-43 &   23.3974609 &     30.708994 & 37.28 & 36.72 &  80 &   0.28 &    H \\
L10-50 &   23.4197102 &     30.709921 & 37.06 & 36.84 & 100 &   0.60 &    S \\
L10-68 &   23.4672909 &     30.942835 & 36.93 & 36.65 & 112 &   0.52 &    S \\
L10-79 &   23.4922905 &     30.810335 & 37.14 & 36.61 &  80 &   0.29 &    H \\
L10-98 &   23.5528698 &     30.586670 & 36.61 & 36.10 &  62 &   0.31 &    H \\
L10-122 &   23.6327095 &     30.944935 & 36.94 & 36.87 & 126 &   0.85 &    S \\
L10-131 &   23.6745396 &     30.626440 & 36.87 & 36.52 & 156 &   0.45 &    S \\
L10-132 &   23.6857471 &     30.710779 & 36.39 & 35.86 &  50 &   0.29 &    H \\
L10-133 &   23.7286701 &     30.688061 & 36.94 & 36.48 &  72 &   0.35 &    H \\
L10-136 &   23.7550793 &     30.638155 & 36.65 & 36.27 & 124 &   0.42 &    S \\
L10-137 &   23.7632504 &     30.619329 & 36.61 & 36.53 & 124 &   0.83 &    S \\
\enddata
\tablenotetext{a}{\small{Measured in the \ha image.}}
\tablenotetext{b}{\small{Derived using $L = 4\pi \rm {d}^{2}\times$ flux adopting d = 800 kpc.}}
\tablenotetext{c}{\small{Size derived using 1\arcsec = 3.88 pc.}}
\tablenotetext{d}{\small{H:\h2 with \rat $<$ 0.4, and blue stars inside;    
S: Superbubbles (Larger than $D =$ 100 pc and a number of blue stars inside).}}
\tablenotetext{e}{\small{An oxygen-dominated SNR with little emission in \ha and \s2 (L10).}}
\end{deluxetable}

\begin{deluxetable}{cccccccc}
\tablecaption{Characteristics of M33 SNR Candidates of Different Morphological Categories
              \label{table2}}
\tablehead{\colhead{Type}  & \colhead{Number\tablenotemark{a}} &
           \colhead{$L$(\hab)} & \colhead{$L$(\s2)} & \colhead{$D$} &
           \colhead{\footnotesize{\ratb}}   & \colhead{Environment} &            
           \colhead{Description}}
\tabletypesize{\small}
\tablewidth{0pt}
\rotate\startdata
A1 & 41(5)  & Moderate & Moderate & Moderate & High & Isolated & \footnotesize{Complete shells}                    \\
A2 & 19(0)  & High     & High     & Small    & High & Isolated & \footnotesize{Center-bright remnants} \\   
A3 & 29(9)  & Low      & Low      & Small    & Low  & Isolated & \footnotesize{Diffuse and extended shells}        \\ 
\hline             
B1 & 36(3)  & Moderate & Moderate & Moderate & High & Isolated & \footnotesize{Partial shells}                     \\
B2 & 24(0)  & High     & High     & Moderate & Low  & Confused & \footnotesize{Bright partial shells}              \\
B3 & 28(11) & Low      & Moderate & Large    & High & Isolated & \footnotesize{Diffuse partial shells}             \\
\hline
C\tablenotemark{b}  & 21(1) & ...      & ...      & ...   & ...  & ...    &  ...                               \\
\enddata 
\tablenotetext{a}{\small{Values in parentheses represent numbers of Type Ia remnants.}}
\tablenotetext{b}{\small{Ambiguous objects, excluding A-class and B-class remnants.}}
\end{deluxetable}

\clearpage

\begin{deluxetable}{cccccccccccc}
\tabletypesize{\small}
\tablecaption{A List of M33 SNR Candidates \label{table3}}
\tablewidth{0pt}
\tablehead{
\colhead{ID} & \colhead{\footnotesize{R.A.(J2000.0)\tablenotemark{a}}} &
\colhead{\footnotesize{Decl.(J2000.0)\tablenotemark{a}}} &
\colhead{\footnotesize{log $L$(\hab)\tablenotemark{b}}}    &
\colhead{\footnotesize{log $L$(\s2)\tablenotemark{b}}}     &
\colhead{$D$\tablenotemark{c}}   &   \colhead{\footnotesize{\ratb}}          &
\colhead{\footnotesize{Morphology}}  &  \colhead{$N$(OB)\tablenotemark{d}}  & \colhead{\footnotesize{Progenitor}} & \colhead{\footnotesize{Comments}}   \\
   & \colhead{\footnotesize{(Degree)}}    & \colhead{\footnotesize{(Degree)}}        & 
\colhead{\footnotesize{(erg $s^{-1}$)}}   & \colhead{\footnotesize{(erg $s^{-1}$)}}  &
\colhead{(pc)} &   & \colhead{\footnotesize{Type}}   &  & \colhead{\footnotesize{Type}}  &  & }
\rotate\startdata
    1 &   23.1074047 &     30.501122    & 36.16  & 36.03 &  81    &   0.74  &     B3  &     4  &   CC  &      \\
    2 &   23.1160583 &     30.595718    & 36.05  & 36.04 &  72    &   0.98  &     B3  &     6  &   CC  &      \\
    3 &   23.1306095 &     30.592470    & 36.02  & 35.79 &  28    &   0.59  &      C  &    18  &   CC  &     L10-2 \\
    4 &   23.1473484 &     30.588837    & 36.36  & 36.06 &  85    &   0.50  &     A3  &     2  &   CC  &      \\
    5 &   23.1548519 &     30.298412    & 36.15  & 35.93 &  85    &   0.60  &     B3  &     0  &   Ia  &      \\
    6 &   23.1657581 &     30.465267    & 35.36  & 35.33 &  36    &   0.93  &     B3  &     3  &   CC  &      \\
    7 &   23.1676331 &     30.272575    & 35.65  & 35.37 &  44    &   0.52  &     A3  &     1  &   Ia  &      \\
    8 &   23.1694889 &     30.275404    & 35.36  & 35.26 &  42    &   0.79  &     A3  &     1  &   Ia  &      \\
    9 &   23.1705875 &     30.530849    & 36.36  & 36.07 &  86    &   0.51  &     A3  &     1  &   Ia  &      \\
   10 &   23.1779709 &     30.605574    & 36.07  & 35.76 &  58    &   0.49  &      B  &    17  &   CC  &      \\
\enddata
\tablenotetext{a}{\small{Measured in the \ha image.}}
\tablenotetext{b}{\small{Derived using $L = 4\pi \rm {d}^{2}\times$ flux adopting d = 800 kpc.}}
\tablenotetext{c}{\small{Size derived using 1\arcsec = 3.88 pc.}}
\tablenotetext{d}{\small{Numbers of OB stars within 100 pc from the center of individual remnants.} \\
(This table is available in its entirety in a machine-readable form in the online journal.\\ 
 A portion is shown here for guidance regarding its form and content.)}
\end{deluxetable}

\clearpage

\begin{deluxetable}{ccccccc}
\tablecaption{Power Law Indices for Cumulative Size Distributions of SNR Candidates in Nearby Galaxies \label{table4}}        
\tablehead{\colhead{Galaxies} & \colhead{$N$}   & \colhead{Fitting Range of $D$} &
           \colhead{$\alpha$\tablenotemark{*}}  & \colhead{Phase} & \colhead{Reference}  & \colhead{Wavelength}   }
\tabletypesize{\small}
\tablewidth{0pt}
\rotate\startdata
M33 & 111 & 17 $\sim$ 50 pc & 2.38 $\pm$ 0.05 & ST   & This work    & Optical (LGGS)             \\
M33 &  46 & 19 $\sim$ 36 pc & 2.23 $\pm$ 0.08 & ST   & This work    & Optical (LGGS)             \\
M33 &  47 & 13 $\sim$ 33 pc & 2.72 $\pm$ 0.14 & ST   & G98 & Optical                        \\
M33 &  45 & 19 $\sim$ 36 pc & 2.24 $\pm$ 0.13 & ST   & G98 & Optical                        \\
M33 &  18 &  9 $\sim$ 21 pc & 2.82 $\pm$ 0.04 &      ...       & L10 & X-ray, Optical(LGGS)            \\
M33 &  69 & 21 $\sim$ 50 pc & 1.60 $\pm$ 0.03 &      ...       & L10 & X-ray, Optical(LGGS)            \\ 
M33 &  40 & 19 $\sim$ 36 pc & 1.62 $\pm$ 0.07 &      ...       & L10 & X-ray, Optical(LGGS)            \\ 
LMC &  33 & 15 $\sim$ 55 pc & 1.34 $\pm$ 0.04 & Free expansion & B10 & X-ray, Optical, Radio          \\ 
SMC &  11 & 25 $\sim$ 50 pc & 1.18 $\pm$ 0.03 & Free expansion & B10 & X-ray, Optical, Radio          \\
MW  & 111 & 15 $\sim$ 30 pc & 3.60 $\pm$ 0.06 & Radiative      & \citet{pav13} & Radio ($\Sigma - D$ relation)  \\
M31 &  85 & 17 $\sim$ 50 pc & 2.53 $\pm$ 0.04 & ST   & L14 & Optical (LGGS)             \\
\enddata 
\tablenotetext{*}{We derived the power law indices using the catalogs of remnants in the references.}
\end{deluxetable}

\clearpage

\clearpage

\clearpage

\clearpage

\clearpage

\clearpage

\clearpage

\clearpage

\clearpage

\clearpage

\clearpage

\clearpage

\clearpage

\clearpage

\clearpage

\clearpage

\clearpage

\clearpage

\clearpage


\begin{thebibliography}{}

\bibitem[Asvarov(2014)]{asv14} Asvarov, A.~I.\ 2014, \aap, 561, A70 

\bibitem[Badenes et al.(2009)]{bad09} Badenes, C., Harris, J., Zaritsky, D., \& Prieto, J.~L.\ 2009, \apj, 700, 727 

\bibitem[Badenes et al.(2010)]{bad10} Badenes, C., Maoz, D., \& Draine, B.~T.\ 2010, \mnras, 407, 1301 [B10]

\bibitem[Bandiera \& Petruk(2010)]{ban10} Bandiera, R., \& Petruk, O.\ 2010, \aap, 509, A34

\bibitem[Blair \& Long(2004)]{bla04} Blair, W.~P., \& Long, K.~S.\ 2004, \apjs, 155, 101

\bibitem[Blair et al.(2012)]{bla12} Blair, W.~P., Winkler, P.~F., \& Long, K.~S.\ 2012, \apjs, 203, 8 

\bibitem[Blondin et al.(1998)]{blo98} Blondin, J.~M., Wright, E.~B., Borkowski, K.~J., \& Reynolds,
S.~P.\ 1998, \apj, 500, 342 

\bibitem[Chu \& Kennicutt(1988)]{chu88} Chu, Y.-H., \& Kennicutt, R.~C., Jr.\ 1988, \aj, 96, 1874

\bibitem[Dodorico et al.(1978)]{dod78} Dodorico, S., Benvenuti, P., \& Sabbadin, F.\ 1978, \aap, 63, 63 


\bibitem[Dopita et al.(2010)]{dop10} Dopita, M.~A., Blair, W.~P., Long, K.~S., et al.\ 2010, \apj, 710, 964

\bibitem[Dopita \& Sutherland(2003)]{dop03} Dopita, M.~A., \& Sutherland, R.~S.\ 2003,
Astrophysics of the Diffuse Universe (Berlin: Springer) 

\bibitem[Draine(2011)]{dra11} Draine, B.~T.\ 2011, Physics of the Interstellar and Intergalactic Medium 
(Princeton, NJ:Princeton Univ. Press), Chapter 39

\bibitem[Franchetti et al.(2012)]{fra12} Franchetti, N.~A., Gruendl, R.~A., Chu, Y.-H., et al.\ 2012, \aj, 143, 85 

\bibitem[Gordon et al.(1999)]{gor99} Gordon, S.~M., Duric, N., Kirshner, R.~P., 
Goss, W.~M., \& Viallefond, F.\ 1999, \apjs, 120, 247 

\bibitem[Gordon et al.(1998)]{gor98} Gordon, S.~M., Kirshner, R.~P., Long, K.~S., et al.\ 1998, \apjs, 117, 89 [G98]

\bibitem[Green(2009)]{gre09} Green, D.~A.\ 2009, BASI, 37, 45

\bibitem[Green(1984)]{gre84} Green, D.~A.\ 1984, \mnras, 209, 449

\bibitem[Gvaramadze et al.(2010)]{gva10} Gvaramadze, V.~V., Kroupa, P., \& Pflamm-Altenburg, J.\ 2010, \aap, 519, A33 


\bibitem[Hughes \& Helfand(1984)]{hug84} Hughes, J.~P., \& Helfand, D.~J.\ 1984, \baas, 16, 927

\bibitem[Jennings et al.(2012)]{jen12} Jennings, Z.~G., Williams, B.~F., Murphy, J.~W., et al.\ 2012, \apj, 761, 26 

\bibitem[Leonidaki et al.(2013)]{leo13} Leonidaki, I., Boumis, P., \& Zezas, A.\ 2013, \mnras, 429, 189 

\bibitem[Lee et al.(2002)]{lee02} Lee, M.~G., Kim, M.,  Sarajedini, A., Geisler, D., 
\& Gieren, W.\ 2002, \apj, 565, 959 

\bibitem[Lee et al.(2014)]{lee14} Lee, J.~H., \& Lee, M.~G., 2014, \apj, 786, 130 [L14]

\bibitem[Long et al.(1990)]{lon90} Long, K.~S., Blair, W.~P., Kirshner, R.~P., \& Winkler, P.~F.\ 1990, \apjs, 72, 61

\bibitem[Long et al.(2010)]{lon10} Long, K.~S., Blair, W.~P., Winkler, P.~F., et al.\ 2010, \apjs, 187, 495 [L10]

\bibitem[Long et al.(1996)]{lon96} Long, K.~S., Charles, P.~A., Blair, W.~P., \& Gordon, S.~M.\ 1996, \apj, 466, 750 

\bibitem[Magnier et al.(1995)]{mag95} Magnier, E.~A., Prins, S., van Paradijs, J., et al.\ 1995, \aaps, 114, 215

\bibitem[Massey et al.(2007)]{mas07} Massey, P., McNeill, R.~T., Olsen, K.~A.~G., et al.\ 2007, \aj, 134, 2474 

\bibitem[Massey et al.(2006)]{mas06} Massey, P., Olsen, K.~A.~G., Hodge, P.~W., et al.\ 2006, \aj, 131, 2478

\bibitem[Mathewson et al.(1983)]{mat83} Mathewson, D.~S., Ford, V.~L., Dopita, M.~A., et al.\ 1983, \apjs, 51, 345 

\bibitem[Matonick \& Fesen(1997)]{mat97a} Matonick, D.~M., \& Fesen, R.~A.\ 1997, \apjs, 112, 49

\bibitem[Matonick et al.(1997)]{mat97b} Matonick, D.~M., Fesen, R.~A., Blair, W.~P., \& Long, K.~S.\ 1997, \apjs, 113, 333

\bibitem[McKee \& Ostriker(1977)]{mck77} McKee, C.~F., \& Ostriker, J.~P.\ 1977, \apj, 218, 148

\bibitem[Mills et al.(1984)]{mil84} Mills, B.~Y., Turtle, A.~J., Little, A.~G., \& Durdin, J.~M.\ 1984,
AuJPh, 37, 321 


\bibitem[Pannuti et al.(2002)]{pan02} Pannuti, T.~G., Duric, N., Lacey, C.~K., et al.\ 2002, \apj, 565, 966

\bibitem[Pannuti et al.(2007)]{pan07} Pannuti, T.~G., Schlegel, E.~M., \& Lacey, C.~K.\ 2007, \aj, 133, 1361

\bibitem[Paturel et al.(2003)]{pat03} Paturel, G., Petit, C., Prugniel, P., et al.\ 2003, \aap, 412, 45 

\bibitem[Pavlovi{\'c} et al.(2013)]{pav13} Pavlovi{\'c}, M.~Z., Uro{\v s}evi{\'c},
D., Vukoti{\'c}, B., Arbutina, B., \& G{\"o}ker, {\"U}.~D.\ 2013, \apjs, 204, 4 

\bibitem[Regan \& Vogel(1994)]{reg94} Regan, M.~W., \& Vogel, S.~N.\ 1994, \apj, 434, 536 

\bibitem[Schlafly \& Finkbeiner(2011)]{sch11} Schlafly, E.~F., \& Finkbeiner, D.~P.\ 2011, \apj, 737, 103

\bibitem[Sonbas et al.(2009)]{son09} Sonbas, E., Akyuz, A., \& Balman, S.\ 2009, \aap, 493, 1061

\bibitem[Truelove \& McKee(1999)]{tru99} Truelove, J.~K., \& McKee, C.~F.\ 1999, \apjs, 120, 299 

\bibitem[Woltjer(1972)]{wol72} Woltjer, L.\ 1972, \araa, 10, 129

\end{thebibliography}
\end{document}